\documentclass[12pt,draftcls,onecolumn]{IEEEtran} 
\usepackage{mathrsfs, amsmath}
\usepackage{amssymb}
\usepackage{cite}
\usepackage{amsfonts}
\usepackage{graphicx}
\usepackage{color}

\def\rep#1{(\ref{#1})}

\newcommand{\R}{\mathbb{R}}

\def\send#1#2{\stackrel{#1}{\hbox to #2{\rightarrowfill}}}
\def\-{\!\!\!\!\!-}

 \def\qed{ \rule{.1in}{.1in}}
\def\eq#1{\begin{equation}#1\end{equation}}

\newcommand{\rank}{{\rm rank\;}}

\def\scr#1{{\cal #1}}

\newcommand{\matt}[1]{\left[ \begin{matrix} #1 \end{matrix}\right]}

\newcommand{\dfb}{\stackrel{\Delta}{=}}

\newtheorem{theorem}{Theorem}
\newtheorem{lemma}{Lemma}

\newtheorem{proposition}{Proposition}
\newtheorem{corollary}{Corollary}

\def\qed{ \rule{.1in}{.1in}}

\def\R{{\rm I\!R}}

\newcounter{seqn}[equation]
\def\theseqn{\arabic{equation}\alph{seqn}}

\def\endseqn{\eqno \@seqnnum
$$\ignorespaces}
\def\@seqnnum{(\theseqn)}

\newskip\mcentering \mcentering=0pt plus 1000pt minus 1000pt

\def\meqalignno#1{
\halign to\displaywidth{
    \hbox to 0pt{\kern\displaywidth\llap{$##$}\hss}\tabskip=\mcentering
    &\hfil$\displaystyle{##}$\tabskip=\mcentering
   &&$\displaystyle{{}##}$\hfil\tabskip=\mcentering
    \crcr
    #1\crcr}}

\def\rep#1{(\ref{#1})}

\def\eq#1{\begin{equation}#1\end{equation}}

\def\dspace{\multiply\normalbaselineskip 150
		  \divide\normalbaselineskip 100 \normalbaselines
		  \csname @@normalbaselineskip\endcsname\normalbaselineskip}
\def\sspace{\multiply\normalbaselineskip 200
		 \divide\normalbaselineskip 300 \normalbaselines
		 \csname @@normalbaselineskip\endcsname\normalbaselineskip}
\def\sdspace{\multiply\normalbaselineskip 160
		 \divide\normalbaselineskip 150 \normalbaselines
		 \csname @@normalbaselineskip\endcsname\normalbaselineskip}

\def\@{\tilde}

\def\3dot#1{\buildrel\textstyle...\over#1}

\begin{document}
\title{Undirected Rigid Formations are Problematic\thanks{The authors thank  Walter Whiteley, York University, Toronto, Canada for several useful discussions which have contributed to this work.}}
\author{S. Mou \hspace{.1in} A. S. Morse \hspace{.1in}  M. A. Belabbas  \hspace{.1in}
 Z. Sun \hspace{.1in} B. D. O. Anderson
\thanks{S. Mou and A. S. Morse are with the Department of Electrical Engineering, Yale University.  M. A. Belabbas is with Department of Electrical and Computer Engineering, University of Illinois at Urbana-Champaign. Z. Sun and B. D. O. Anderson are with NICTA \& Australian National University. {\tt\small shaoshuai.mou@yale.edu,  as.morse@yale.edu,  belabbas@illinois.edu, zhiyong.sun@anu.edu.au,  brian.anderson@anu.edu.au}. Corresponding author: S. Mou. }
\thanks{The research of S. Mou and  A. S. Morse was supported by  the
US Air Force Office of Scientific Research   and the by National Science Foundation.
The research of A. Belabbas  was supported in part by the Army Research Office under
PECASE Award W911NF-091-0555 and by the Office of Naval Research
under MURI Award 58153-MA-MUR. B. D. O Anderson's research was  is supported by Australian Research Council's
Discovery
 Project DP-110100538 and  National ICT Australia-NICTA.}
}

\markboth{ IEEE Transactions on Automatic Control, Accepted. }{Shell \MakeLowercase{\textit{et al.}}: Bare Demo of
IEEEtran.cls for Journals}

\maketitle
\begin{abstract} By an {\em undirected rigid formation} of mobile autonomous agents is meant a formation based
on graph rigidity   in which each
pair of  ``neighboring'' agents
is  responsible for maintaining a prescribed target distance   between them.
In  a recent paper  a systematic method was proposed for devising  gradient control laws
 for  asymptotically stabilizing a large class of  rigid, undirected formations in two-dimensional
  space assuming all agents are described by kinematic point models.
  The aim of this paper is to explain what happens
  to such formations  if neighboring agents  have
 slightly different
 understandings of what the desired  distance  between them is supposed to be  or equivalently  if neighboring agents  have differing  estimates
of  what the  actual distance between them is.
 In either case, what one would expect
 would be  a gradual distortion
of the  formation  from its target shape as discrepancies in desired  or sensed distances  increase.
While this is observed for the gradient laws  in question, something else quite unexpected
 happens at the same time. It is shown  that
    for
  any rigidity-based,
 undirected  formation of this type which is  comprised of three or more agents,
  that if some neighboring agents
 have slightly different understandings  of what the desired
  distances between them are suppose to be, then almost for certain, the trajectory of the resulting
    distorted  but rigid formation will converge exponentially fast to a closed circular orbit in two-dimensional space
    which is traversed
    periodically at a constant angular speed.\end{abstract}


\IEEEpeerreviewmaketitle

\section{Introduction}

 The problem of coordinating a large   network  of  mobile autonomous  agents
by means of distributed control has raised a number of issues
concerned with  the forming,  maintenance
 and real-time modification of  multi-agent networks of all types.
One of the most natural and useful tasks along these lines is to organize a network of agents into an
application-specific ``formation'' which might be used for such tasks as  environmental monitoring, search,
 or simply moving
the agents efficiently from one location to another.
By a multi-agent  formation   is usually meant a collection of agents  in  real two
 or three  dimensional
space whose inter-agent  distances are all essentially constant
over  time, at least
under ideal  conditions. One approach to maintaining such formations is based on the idea of
``graph rigidity'' \cite{asi.rothII,rigid}.
   Rigid formations can be ``directed'' \cite{johnbs,CBSB09SIAM,JBJV07IJR},
 ``undirected'' \cite{krickb,RR02IFAC}, or some combination of the two. The appeal of the rigidity based approach  is that it has the potential for providing  control laws which are totally distributed in that the only information which each agent needs to sense is the relative positions of its nearby neighbors. 

By an {\em undirected rigid formation} of mobile autonomous agents is meant a formation based
on graph rigidity   in which each
pair of  ``neighboring'' agents  $i$ and $ j$
are  responsible for maintaining the prescribed target distance  $d_{ij}$ between them.
In  \cite{krickb}  a systematic method was proposed for devising  gradient control laws
 for  asymptotically stabilizing a large class of  rigid, undirected formations in two-dimensional
  space assuming all agents are described by kinematic point models. This particular methodology is perhaps the most comprehensive currently in existence for maintaining  formations based on graph rigidity. In \cite{robust} an effort was made to understand what happens
  to such formations  if neighboring agents $i$ and $j$ have
 slightly different
 understandings of what the desired  distance $d_{ij}$  between them is supposed to be.
The question is relevant because no two positioning controls
can be expected to  move agents to  precisely specified positions because of inevitable imprecision
in the physical comparators used  to  compute the   positioning errors. The question is also relevant
because it is  mathematically
equivalent to determining  what happens if neighboring agents $i$ and $j$ have differing  estimates
of  what the  actual distance between them is.
In either case, what one might expect
 would be  a gradual distortion
of the  formation  from its target shape as discrepancies in desired  or sensed distances  increase.
While this is observed for the gradient laws  in question, something else quite unexpected
 happens at the same time. In particular 
  it turns out  for
  any rigidity-based,
 undirected  formation of the type considered in \cite{krickb} which is  comprised of three or more agents,
  that if some neighboring agents
 have slightly different understandings  of what the desired
  distances between them are suppose to be, then almost for certain, the trajectory of the resulting
    distorted  but rigid formation will converge exponentially fast to a closed circular orbit in $\R^2$
    which is traversed
    periodically at a constant angular speed.
   In \cite{robust}  this was shown to be so for the special
    case of a three agent triangular formation.
    The aim of this paper is to explain why this same  phenomenon
         also occurs with any undirected rigid formation in the plane consisting of three or more agents.


\subsection{Organization}

This paper is organized as follows.  In \S \ref{GR} we briefly summarize the concepts from graph rigidity theory which are used in this paper. In \S \ref{uf} we describe the undirected rigidity-based control law introduced in \cite{krickb} and we develop a model, called the ``overall system,'' which exhibits the kind of mismatch error $\mu$ we intend to study.  In \S \ref{esystem} we develop and discuss in detail, a separate  self - contained ``error system'' $\dot{\epsilon} = g(\epsilon,\mu)$  whose existence is crucial to understanding the effect of mismatch errors.
The error system  can only be
defined locally and its existence  is not obvious. Theorem \ref{china11}  states that
overall system's error $e$  satisfies the error system's dynamics along trajectories of the overall
system which lie within a suitably defined open subset $\scr{A}$.
The theorem is proved
in \S \ref{esystemd} by appealing to the inverse function theorem.
In \S \ref{exun} we prove that the ``unperturbed'' error system   $\dot{\epsilon} = g(\epsilon,0)$  is locally
exponentially stable 
 We then exploit the well know robustness of exponentially stable dynamical systems to prove in \S \ref{expert} that even with a mismatch error, the error system remains locally exponentially stable
  provided the norm of the mismatch error  $\mu$ is sufficiently small.
  Finally in
\S \ref{expos} we show that if   the overall system  starts
 in a state within $\scr{A}$
  at which the overall system's error $e$ is sufficiently close
to  the  output $\mathfrak{e}_{\mu}$ of the  error system assuming the error system is in equilibrium, then the state
of the overall system remains within  $\scr{A}$ for all time and its error $e$ converges exponentially fast to  $\mathfrak{e}_{\mu}$.

In \S \ref{redu} we develop a special  $2\times 2$
``square subsystem''  whose behavior along trajectories of the overall system enables us to predict the behavior of the overall system. An  especially important property of this subsystem is that it is linear along trajectories of the overall system
for which the error system's output is constant.

In \S \ref{anals} we turn to the analysis  of the overall system which we carry out in two steps. First, in Section
\ref{step1} we  consider  the situation when the  overall system error $e$ has already converged to
$\mathfrak{e}_{\mu}$.
In \S \ref{mme} we develop conditions on the mismatch error $\mu$ under which
 the state of the overall system will be nonconstant, even though  $e$ is constant.
In \S \ref{horns} we characterize  the behavior of trajectories  of the overall system assuming  $e$  is constant. The main result in the section, stated in Theorem
 \ref{main2}, is that for a large class of formations,  the type of mismatch error we are considering will almost certainly cause the formation to rotate  at a constant angular speed about a fixed point in two-dimension space, provided the norm of the mismatch error is sufficiently small.

 The second step in the analysis is carried out in Section  \S \ref{NEA}. The main result of this paper,  Theorem \ref{Maino}, states that if a formation starts out in a state in $\scr{A}$ at which its error
 $e$  is equal to a value of the error systems's output for which  the error system's state is
 in the domain of  attraction of the error system's equilibrium,  then the formation's state will converge exponentially fast to the state of a formation moving at constant angular speed in a circular orbit
 in the plane.

 \subsection{Graph Rigidity} \label{GR}

The aim of this section is to briefly summarize  the concepts from graph rigidity theory which will be used in this paper.
By a  {\em framework} in $\R^2$    is meant a set of $n\geq 3$ points in the real plane with coordinate vectors $x_i,\;i\in\mathbf{n} \dfb \{1,2,\ldots,n\}$, in $\R^2$ together with  a simple, undirected  graph $\mathbb{G}$ with $n$ vertices labeled $1,2,\ldots, n$ and $m$ edges labeled $1,2,\ldots ,m$.
 We denote such a framework by the pair $\{\mathbb{G},x\}$ where $x$ is the {\em multi-point} $x=\matt{x_1' &x_2'&\ldots &x_n'}'$.
An important property of any framework is that its shape does not change under  ``translations'' and ``rotations.''
To make precise what is meant by this let us agree to say that
    a {\em translation}   of a multi-point $x = \matt{x_1' &x_2' &\cdots & x_n'}'$ is a function of the form
$\matt{x_1'&x_2'&\cdots &x_n'}'\longmapsto \matt{x'_1+y'&x'_2+y' & \cdots & x_n'+y'}'$ where $y$ is a vector
in $\R^2$. Similarly,
a  {\em rotation} of  a multi-point $x$ is a function of the form
 $\matt{x'_1 & x'_2 & \cdots & x'_n}'\longmapsto \matt{(Tx_1)' & (Tx_2)' & \cdots & (Tx_n)'}'$ where
 $T:\R^2\rightarrow\R^2$ is
 a rotation matrix. The set of all such translations and rotations
 together with composition forms a transformation group which we denote by $\mathfrak{G}$; this group is
 isomorphic to the   special Euclidean group $SE(2)$. By the {\em orbit} of $x\in\R^{2n}$, written
$\mathfrak{G}x$ is meant the set $\{\gamma(x):\gamma\in\mathfrak{G}\}$. Correspondingly, the {\em orbit of a framework} $\{\mathbb{G},x\}$ is the set of all frameworks $\{\mathbb{G},y\}$ for which $y$ is in the orbit of $x$.
   By $\mathbb{G}$'s {\em edge function} $\phi:\R^{2n}\rightarrow \R^m$    is meant the map
  $x\longmapsto \matt{||x_{i_1}-x_{j_1}||^2 &||x_{i_2}-x_{j_2}||^2 &\cdots  & ||x_{i_m}-x_{j_m}||^2}'$, where  for $k\in\mathbf{m}\dfb \{1,2,\ldots,m\}$,
  $(i_k,j_k)$ is the $k$th edge in $\mathbb{G}$. A framework $\{\mathbb{G},x\}$ is a {\em realization} of a non-negative vector $v\in\R^m$  if $\phi(x) = v$; of course not every such vector is realizable. Two frameworks $\{\mathbb{G},x\}$ and $\{\mathbb{G},y\}$ in $\R^2$ are {\em equivalent}  if  they have the same
edge lengths; i.e., if $\phi(x) = \phi(y)$.
 $\{\mathbb{G},x\}$ and $\{\mathbb{G},y\}$ are {\em congruent} if for
each pair of distinct labels  $i,j\in\mathbf{n}$, $||x_i-x_j|| = ||y_i-y_j||$. It is important to recognize that  while
two formations $\{\mathbb{G},x\}$ and $\{\mathbb{G},y\}$   in the same orbit must be congruent, the converse is not necessarily true, even if both formations are ``rigid.''.
Roughly speaking, a framework is rigid if it is impossible to `deform' it by moving its points slightly while holding all of its edge lengths constant. More precisely, a framework  $\{\mathbb{G},x\}$ in $\R^2$ is {\em rigid} if it is congruent to every equivalent
framework $\{\mathbb{G},y\}$ for which $||x-y||$ is sufficiently small. The notion of a rigid framework goes back  several hundred years and has names such as Maxwell, Cayley, and Euler associated with it. In addition to its
use in the study of mechanical structures, rigidity has proved useful
in  molecular biology and in the formulation and solution of  sensor network localization problems \cite{local}. Its application to formation control was originally proposed in  \cite{rigid}.  Unfortunately, it is difficult to completely characterize a rigid framework because of many special cases which defy simple analytical descriptions.  The situation improves if
one  restricts attention to  frameworks for which  the positions of the points are algebraically independent over the rationals. Such frameworks are called {\em  generic} and  their rigidity is  completely characterized by  the so-called {\em rigidity matrix}
$\mathfrak{R}_{m\times 2n}(x) = \frac{\partial \phi(x)}{2\partial x}$.
 The  rigidity matrix appears in the expression for the derivative of the edge function
$\phi(x(t))$ along smooth trajectories
$x(t),\;t\geq 0$; i.e., $\dot{\phi}(x(t)) = 2\mathfrak{R}(x(t))\dot{x}(t)$. It is known that the kernel of
$\mathfrak{R}(x)$ must be a subspace of dimension of at least $3$ \cite{asi.rothII};   equivalently, for all $x$,
$\rank \mathfrak{R}(x) \leq 2n-3$. A framework $\{\mathbb{G},x\}$ is said to be {\em infinitesimally rigid} if $\rank \mathfrak{R}(x) = 2n-3$. Infinitesimally
rigid frameworks are known to be  rigid \cite{asi.rothII,Jackson07noteson}, but examples show that the converse is not necessarily true. However, generic  frameworks  are rigid if and only if they are infinitesimally rigid \cite{asi.rothII}. Any graph $\mathbb{G}$  for which there exists a multi-point $x$ for which
$\{\mathbb{G},x\}$ is a generically rigid framework, is called  a  {\em rigid graph}. Such graphs are completely characterized by  Laman's Theorem \cite{laman} which provides a combinatoric test for graph rigidity. It is obvious that if $\{\mathbb{G},x\}$ is infinitesimally rigid, then so is any
other framework in the same orbit.

An infinitesimally rigid framework is {\em minimally infinitesimally rigid}  if it is infinitesimally rigid
and if the removal of an edge in the framework causes the framework to lose rigidity. It is known that an  infinitesimally rigid framework is minimally infinitesimally rigid if and only if $m=2n-3$ \cite{asi.rothII,RD05Compu}.  An infinitesimally rigid framework $\{\mathbb{G},x\}$ can be ``reduced''
to a minimally infinitesimally rigid framework $\{\tilde{\mathbb{G}},x\}$, with $\tilde{\mathbb{G}}$ a spanning subgraph of $\mathbb{G}$, by simply removing ``redundant'' edges from $\mathbb{G}$. Equivalently, $\{\mathbb{G},x\}$ can be reduced
to a minimally infinitesimally rigid framework $\{\tilde{\mathbb{G}},x\}$ by deleting the  linearly dependent rows from the rigidity matrix $\mathfrak{R}(x)$,  and then deleting the corresponding edges from $\mathbb{G}$ to obtain $\tilde{\mathbb{G}}$. The rigidity matrix of $\{\tilde{\mathbb{G}},x\}$, namely $\tilde{\mathfrak{R}}(x)$, is related to the $\mathfrak{R}(x)$ by an equation of the form $\tilde{\mathfrak{R}}(x) = \tilde{P} \mathfrak{R}(x)$ for a suitably defined matrix $\tilde{P} $ of ones and zeros. 

 In this paper we will  call a framework a {\em formation}.  We will deal exclusively with formations which are
  infinitesimally rigid.

\section{Undirected Formations}\label{uf}

We consider a formation in the plane consisting of $n\geq 3$ mobile autonomous agents \{eg, robots\}
 labeled $1,2,\ldots,n$. We assume the desired formation is specified in part,  by a  graph $\mathbb{G}$
 with $n$ vertices labeled  $1,2,\ldots,n$
and $m$ edges labeled $1,2,\ldots,m$.
 We write $k_{ij}$ for the label
  of that  edge which
connects adjacent vertices $i$ and $j$.
 Thus $k_{ij} = k_{ji}$. We call agent $j$ a {\em neighbor} of agent $i$ if
 vertex $j$ is adjacent to vertex $i$ and we write
 $\scr{N}_i$ for the labels of agent $i$'s neighbors.

We assume that the desired {\em target distance} between agent $i$ and neighbor $j$ is $d_{ij}$ where $d_{ij}$
 is a positive number. We assume that agent $i$  is tasked with the job of maintaining the specified target
  distances to each of its neighbors. However unlike \cite{krickb} we do not assume that the target distances
$d_{ij}$  and  $d_{ji}$ are necessarily equal.
 Instead  we assume that
$|d_{ji} - d_{ij}| \leq\beta_{k_{ij}}$ 
were $\beta_{k_{ij}}$ is a small nonnegative number  bounding the discrepancy in the two agents understanding of
 what  the  desired distance  between them is suppose to be. We assume that in the unperturbed case when there is no
 discrepancy between $d_{ij}$ and $d_{ji}$, these distances are realizable
 by a specific set of points in the plane with coordinate vectors $y_1,y_2,\ldots,y_n$ such  that
 at the  multi-point  $y= \matt{y_1' &y_2' &\cdots y_n'}'$,  the resulting
 formation  $\{\mathbb{G}, y\}$ is  infinitesimally rigid. We call
 $\{\mathbb{G}, y\}$ as well as all formations in its orbit, {\em target formations}.

 In this paper we will assume
 that  any formation  $\{\mathbb{G}, x\}$ which is equivalent to target formation  $\{\mathbb{G}, y\}$, is infinitesimally rigid. While this  is not necessarily true for every possible set
 of realizable target distances, it is true generically, for almost every such  set. This is a consequence of Theorem 5.5 of \cite{hendrickson}. An implication of this assumption is that
the set of  all formations equivalent to target formation  $\{\mathbb{G},y\}$
is equal to the {\em finite} union of a set of disjoint orbits  \cite{nembeddings}. We will assume that there are $n_o >0 $ such orbits, that $\{\mathbb{G},y^i\}$  is a representative of orbit $i$, and that $\{\mathbb{G},y^1\}$ is the target formation $\{\mathbb{G},y\}$.

We assume that agent $i$'s motion is described in global coordinates by the simple
kinematic point model
\eq{\dot{x}_i = u_i,\;\;\;\;i\in\mathbf{n}. \label{int}}
We further assume that for $i\in\mathbf{n}$, agent $i$ can 
 measure the relative position
$x_j-x_i$ of each of its neighbors $j\in\scr{N}_i$. The aim of the formation control problem posed in \cite{krickb} is to devise individual agent controls  which, with $x=\matt{x_1' &x_2' &\cdots x_n'}'$, will  cause the resulting formation
$\{\mathbb{G},x\}$ to approach a target formation and come to rest as $t\rightarrow \infty$.
The control law for agent $i$ proposed in  \cite{krickb} to accomplish this   is
$$u_i= \sum_{j\in\scr{N}_i}(x_j-x_i)(||x_j-x_i||^2-d_{ij}^2).$$
Application of such controls 
  to the agent models \rep{int}  yields the equations
\eq{\dot{x}_i =\sum_{j\in\scr{N}_i}(x_j-x_i)(||x_j-x_i||^2-d_{ij}^2),\;\;\;\;\;i\in\mathbf{n}\label{int2}}
Our aim is to express these equations in state space form. To do this it is convenient
 to   assume that
each edge in $\mathbb{G}$
is ``oriented'' with  a specific direction,  one end of the edge being its `head' and the other
being its `tail.' To proceed,
  let us write $H_{m\times n}$ for that matrix whose $ki$th entry is $h_{ki} = 1$ if vertex $i$ is the head of
oriented edge $k$,
$h_{ki} = -1$ if vertex $i$ is the tail of oriented edge $k$ and $h_{ki}=0$ otherwise. Thus $H$ is a matrix
of $1$s, $-1$s and $0$s with exactly one $1$ and one $-1$ in each row. Note that $H$ is the  transpose of
 the incidence matrix of the oriented graph $\mathbb{G}$;  because $\mathbb{G}$ is connected, the
 rank of $H$
 is $n-1$ \cite{NDeo74Pren}.   Next define for each edge $k_{ij}$,
\eq{z_{k_{ij}} = \chi_{ij}(x_i-x_j  )\label{pi}}
where $\chi_{ij} = 1$ if $i$ is the head of edge $k_{ij}$ or  $\chi_{ij} = -1$
 if $i$ is the tail of edge $k_{ij}$.
 The definition of $H$ implies that
\eq{z =\bar{H}x\label{z}}
where 
 $z=\matt{z'_1 &z'_2 &\cdots z'_m}'$,
 $\bar{H}_{2m\times 2n} = H\otimes I_{2\times 2}$,
 $I_{2\times 2}$ is the $2\times 2$ identity and $\otimes $ is the Kronecker product.

Next define $d_{k_{ij}} = d_{ij}$  and $\mu_{k_{ij}} = d_{ij}^2-d_{ji}^2$ for all adjacent vertex pairs $(i,j)$ for which $i$ is the head of edge $k_{ij}$;
clearly $$d_{ij}^2 = d_{k_{ij}}^2\hspace{.5in} {\rm and}\hspace{.5in} d^2_{ji} = d^2_{k_{ij}} -\mu_{k_{ij}}$$ for all such pairs.
Let $e_k:\R^m\rightarrow \R$ denote the $kth$ {\em error function}
\eq{e_{k}(z) = ||z_k||^2 - d_k^2,
\;\;\;\;\;k\in\mathbf{m}.\label{e}}
Write  $\scr{N}^+_i$ for the set of   all $j\in\scr{N}_i$ for which  vertex $i$ is a
head of oriented edge $k_{ij}$. Let
$\scr{N}^-_i$ denote the complement of $\scr{N}^+_i$ in $\scr{N}_i$.
With the $z_{k_{ij}}$  and $z$ as defined in \rep{pi} and \rep{z} respectively, the system of equations
given in \rep{int2}
can  be written as
\eq{\dot{x}_i =  -\sum_{j\in\scr{N}_i^+}z_{k_{ij}}e_{k_{ij}}(z)
 + \sum_{j\in\scr{N}_i^-}z_{k_{ij}}(e_{k_{ij}}(z)+\mu_{k_{ij}}),\;\;\;\;i\in\mathbf{n}. \label{ov}}
These equations in turn can be written compactly in the  form
\eq{\dot{x} = -R'(z)e(z) +S'(z)\mu\label{overall}}
where
 $\mu$ is the  {\em mismatch error}  $\;\mu = \matt{\mu_1 & \mu_2 & \cdots \mu_m}'$,
 $e(z) = \matt{e_1(z) &e_2(z) &\cdots &e_m(z)}'$,
 $R_{m\times 2n}(z) = D'(z)\bar{H}$, $S_{m\times 2n}(z) =D'(z) \bar{J}$,
  $D_{2m\times m}(z) = {\rm diagonal}\{z_1,z_2,\ldots,z_m\}$,
  and $\bar{J}_{2m\times 2n}$ is what
    results when the negative elements in $-\bar{H}$ are replaced by zeros.
It is easy to verify that $R(z)|_{z=\bar{H}x}$ is the rigidity matrix  $\mathfrak{R}(x)$ for the formation $\{\mathbb{G},x\}$ \cite{asi.rothII}.
Note that
because of \rep{z},
\rep{overall} is a smooth self-contained dynamical system of the form $\dot{x} = f(x,\mu)$.
 We shall refer to \rep{overall}  \{with $z=\bar{H}x$\} as the {\em overall system}.

\begin{figure}[h]\centerline{
 \includegraphics[height = 1.5in]{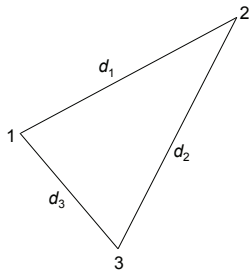}}
    \caption{Undirected Triangular Formation}
    \label{tr1}
   \end{figure}

\noindent{ \bf Triangle Example:} For the  triangle shown in Figure \ref{tr1}
$$u_i = -z_ie_i+z_{[[i]]}(e_{[[i]]}+\mu_{[[i]]}),\;i\in\{1,2,3\}$$
where $[1]=2$, $[2]=3$, $[3] = 1$ and
for $i\in\{1,2,3\}$,
$z_i = x_i-x_{[i]},$ 
 and  $e_i = ||z_i||^2-d_i^2$. 
Application of these
 perturbed controls to  \rep{int} then yields the
 equations
\eq{\dot{x}_i = -z_ie_i +z_{[[i]]}e_{[[i]]}+z_{[[i]]}\mu_{[[i]]},\;\;\;\;\;i\in\{1,2,3\}.\label{epdates0}}

 \section{Error System}\label{esystem}

 Our aim is to study the geometry of the overall system.  Towards this end, first note that
   \eq{\dot{z} = -\bar{H}R'(z)e(z) +\bar{H}S'(z)\mu\label{z2}}
 because of   \rep{z} and \rep{overall}.  This equation and the definitions of  the $e_k$ in \rep{e}
  enable one to write
\eq{\dot{e} = -2R(z)R'(z)e + 2R(z)S'(z)\mu.\label{error1}}
If the target formation $\{\mathbb{G},y\}$ is only infinitesimally rigid but not {\em minimally} infinitesimally rigid there are further constraints imposed on $e$ stemming from the fact there are geometric dependencies  between its components.  Because of this, along  trajectories where $\{\mathbb{G}, x(t)\}$ is infinitesimally rigid, $e$ evolves in a closed proper subset $\scr{E}\subset\R^m$ containing $0$.
We now explain what these dependencies are and in the process define $\scr{E}$.

Set $\tilde{m} = 2n-3$, the rank of the rigidity matrix of $\{\mathbb{G},y\}$.
Suppose  that  $\{\mathbb{G},y\}$  is not minimally infinitesimally rigid in which case  $m > \tilde{m}$.  Let
$\tilde{\mathbb{G}}$ be any spanning subgraph of $\mathbb{G}$ for which $\{\tilde{\mathbb{G}},y\}$ is minimally
infinitesimally rigid.
Write $\tilde{e}$ for the sub-vector of $e$ whose $\tilde{m}$ entries are those entries
in $e$ corresponding to the edges in $\tilde{\mathbb{G}}$. Similarly, write $\widehat{e}$
for those entries in $e$ corresponding to the $m-\tilde{m}$  edges in $\mathbb{G}$ which have been deleted to form $\tilde{\mathbb{G}}$. Let $\tilde{P}$ and $\widehat{P}$ be those matrices for which $\tilde{e} = \tilde{P}e$
and $\widehat{e} = \widehat{P}e$ respectively.  Note that $\matt{\tilde{P}' & \widehat{P}'}$ is a permutation matrix;  therefore  $\tilde{P}\tilde{P}' = I$, $\widehat{P}\widehat{P}' = I$, $\tilde{P}\widehat{P}' = 0$, and
$e = \tilde{P}'\tilde{e} +\widehat{P}'\widehat{e}$.

Recall that each entry $e_s$  in $e$  is, by definition, a function of the form
$(x_i-x_j)'(x_i-x_j) - d_s^2$ where $(i,j)$ is the $s$th edge in $\mathbb{G}$.  The following proposition implies that each such $e_s$ can be expressed  as a  smooth function of $\tilde{e}$ at points  $x$  in a suitably defined open subset of $\R^{2n}$  where $\{\tilde{\mathbb{G}}, x\}$ is minimally infinitesimally rigid.

\begin{proposition}  Let $\{\mathbb{G},y\}$ be a target formation.  There exists an open subset $\scr{A}\subset\R^{2n}$ containing $y$ for which the following statements are true. For each four distinct integers $i,j,k,l$  in $\mathbf{n}$,
 there exists a  smooth function $\eta_{ijkl}:\tilde{P}e(\bar{H}\scr{A})\rightarrow\R$ for which
\eq{(x_i-x_j)'(x_k-x_l ) = \eta_{ijkl}(\tilde{P}e(\bar{H}x)),\;\;\;x\in\scr{A}.
\label{quad2}}
Moreover, $\scr{A}$ is invariant under the action of $\mathfrak{G}$ and  for each $x\in\scr{A}$,
the reduced formation $\{\tilde{G},x\}$, is minimally infinitesimally rigid.
 \label{quadlemma2}\end{proposition}
The proof of this proposition  will be given at the end of this section.

 In view of Proposition \ref{quadlemma2},
   there must be a smooth function $\psi:\tilde{P}e(\bar{H}\scr{A})\rightarrow \R^{(m-\tilde{m})}$ such that
$\widehat{e}(\bar{H}x) = \psi(
\tilde{P}e(\bar{H}x)),\;\;x\in\scr{A}$. Observe that $\psi(0) = 0$ because
  $\widehat{P}e(\bar{H}y) = \psi(\tilde{P}e(\bar{H}y))$ and $e(\bar{H}y)=0$.

Note that
\eq{e(\bar{H}x) = \tilde{P}'\tilde{e}(\bar{H}x) +\widehat{P}'\psi(\tilde{e}(\bar{H}x)),\;\;\;x\in\scr{A} \label{pizzas}}
 because  $e = \tilde{P}'\tilde{e} +\widehat{P}'\widehat{e}$.  Moreover, since $\widehat{P}\tilde{P}' = 0$, $\widehat{P}\widehat{P}' = I$ and
$\tilde{e} = \tilde{P}e$, it must be true that for  $x\in\scr{A}$, $\widehat{P}e(\bar{H}x) = \psi(\tilde{P}e(\bar{H}x))$.  In other words, for  such values of $x$,  $e(\bar{H}x)$ takes values in the subset $$\scr{E} = \{e: \widehat{P}e-\psi(\tilde{P}e) = 0,\;e\in e(\bar{H}\scr{A})\}.$$  It is easy to see that
  $0\in\scr{E}$.

We claim that   for values of $x(t)\in\scr{A}$,
   the {\em reduced error}
 $\tilde{e} = \tilde{P}e$ satisfies the differential equation
\eq{\dot{\tilde{e}} = - 2\tilde{R}\tilde{R}'\tilde{e} - 2\tilde{R}\tilde{R}'F'(\tilde{e})\psi(\tilde{e})  + 2\tilde{R}S'\mu \label{onion22}}
  where $\tilde{R}(\bar{H}x)$
is the rigidity matrix of the minimally infinitesimally rigid formation $\{\tilde{\mathbb{G}},x\}$
and
\eq{F(\tilde{e}) =\frac{\partial}{\partial \tilde{e}}\psi(\tilde{e}).\label{qoo3}}
To understand why this is so, note first that \rep{error1} and \rep{pizzas}
imply that
\eq{\dot{\tilde{e}} =
 - 2\tilde{P}RR'(\tilde{P}'\tilde{e} +\widehat{P}'\psi(\tilde{e})) + 2\tilde{P}RS'\mu. \label{onion}}
 By definition, the rigidity matrix of $\{\tilde{\mathbb{G}},x\}$ is
 $\tilde{R}(\bar{H}x) =\frac{1}{2}\frac{\partial}{\partial x}\tilde{e}(\bar{H}x)$.
Clearly
   $\tilde{P}\frac{1}{2}\frac{\partial}{\partial x}e(\bar{H}x) =\tilde{P} R(\bar{H}x)$
 so $\tilde{R} =  \tilde {P} R$.
 From this and \rep{onion} it follows that
  \eq{\dot{\tilde{e}} = - 2\tilde{R}\tilde{R}'\tilde{e} - 2\tilde{R}(\widehat{P}R)'\psi(\tilde{e})  + 2\tilde{R}S'\mu \label{onion6}}
But by definition,  $R(\bar{H}x) =\frac{1}{2}\frac{\partial}{\partial x}e(\bar{H}x)$.
From this and
\rep{pizzas}  it follows that  $R = \tilde{P}'\tilde{R}
 + \widehat{P}'F\tilde{R}$
 where $F$ is given by  \rep{qoo3}.  Thus
$\widehat{P}R = F\tilde{R}$ which justifies the claim that \rep{onion22} holds.

The preceding easily extends to the case  when $\{\mathbb{G},y\}$ itself is minimally infinitesimally rigid.  In this case, $\tilde{e} = e$ and
\rep{pizzas} holds with  $\tilde{P} = I$, $\widehat{P} = 0$, and $\psi = 0$, while
$\scr{E} = \R^{m}$.

The proof of Proposition \ref{quadlemma2} relies on several geometric facts.    The ones we need are encompassed by the following two lemmas.
\begin{lemma} Let $v_1,v_2,v_3,v_4$ be four vectors in $\R^s$ where $s$ is any  fixed positive integer. Then
\eq{(v_1-v_2)'(v_3-v_4) = \frac{1}{2} \{||v_3-v_2||^2 +||v_1-v_4||^2 - ||v_3-v_1||^2-||v_2-v_4||^2\}.
\label{quad}}
\label{quadlemma}\end{lemma}
\noindent{\bf Proof of Lemma \ref{quadlemma}:} Note that
\eq{(v_1-v_2)'(v_3-v_4) = (v_1-v_3)'(v_3-v_4) - (v_2-v_3)'(v_3-v_4)\label{pop1}}
because $v_1-v_2 = (v_1-v_3) - (v_2-v_3)$. But
$$(v_1-v_3)'(v_3-v_4) = \frac{1}{2}\{||v_1-v_4||^2-||v_3-v_1||^2-||v_3-v_4||^2\}$$
and
$$(v_2-v_3)'(v_3-v_4) = \frac{1}{2}\{||v_2-v_4||^2-||v_3-v_2||^2-||v_3-v_4||^2\}.$$
From these identities and \rep{pop1} it follows that \rep{quad} is true. $\qed $


  In the sequel we will show that
 for any $x = \matt{x_1'& x_2' & \cdots &x_n'}'$ in a  suitably defined open set  of multi-points
  $x$   for which $\{\tilde{\mathbb{G}},x\}$ is minimally  infinitesimally rigid, it is possible
  to express the squared distances between  each pair of points $x_i,x_j$
  in terms of the reduced error $\tilde{P}e(\bar{H}x)$.
   Since
   infinitesimal rigidity demands among other things that for at least one pair of points $p$ and $q$, $x_p\neq x_q$,  nothing will be lost by excluding
   from consideration at the outset, values of $x$ for which $x_p=x_q$. For simplicity we will assume the vertices are labeled so that $x_1\neq x_2$.  Accordingly, let $\scr{X}$
   denote the set of all $x\in\R^{2n}$ for which $x_1\neq x_2$ and
 write   $\delta:\scr{X}\rightarrow \R^{\frac{n(n-1)}{2}}$ for the
   {\em squared distance function}
 $$x\longmapsto\matt{||x_1-x_2||^2 &||x_1-x_3||^2&\cdots &||x_{n-1}-x_n||^2}'.$$

To avoid unnecessarily cluttered  formulas in the statements and proofs of some of the lemmas which follow,
   we will make use  of the function $\rho:\scr{X}\rightarrow \R^{\tilde{m}}$  defined by
 $x\longmapsto \tilde{P}e(\bar{H}x)$. Note that  $\rho(x)$  and $\tilde{P}e(\bar{H}x)$ have the same value at every point $x\in\scr{X}$,
 although their domains are different; consequently they are  different functions. Note also that
  for any transformation $\gamma $ in the restriction of $\mathfrak{G}$ to $\scr{X}$,
  $\rho\circ\gamma =\rho$.  Thus for any subset $\scr{W}\subset \scr{X}$, $\rho^{-1}(\scr{W})$ is
  $\mathfrak{G}$ invariant.

\begin{lemma} Let $\{\mathbb{G},y\}$ be a target formation.
There is an open set $\scr{A}\subset \scr{X}$ containing $y$ and
 a smooth function
$f:\tilde{P}e(\bar{H}\scr{A})\rightarrow  \R^{\frac{n(n-1)}{2}} $
for which \eq{\delta(x) = f(\tilde{P}e(\bar{H}x)),\;\;x\in\scr{A}.\label{func}}
  Moreover, $\scr{A}$ is invariant under the action of $\mathfrak{G}$ and for all $x\in\scr{A}$, the reduced formation $\{\tilde{\mathbb{G}},x\}$ is minimally infinitesimally rigid.
\label{pn2} \end{lemma}
\noindent{\bf Proof of Lemma \ref{pn2}:}
For each nonzero vector $q=\matt{q_1 &q_2}'$ in $\R^2$, let $T_q$ denote the rotation matrix
 $$
 T_q =\frac{1}{||q||} \matt{q_2 & -q_1\cr q_1 &q_2}.$$
 Note that $T_qq = \matt{0 &||q||}'$ and that the function $ q\longmapsto T_q$ is well-defined and smooth on
 $\R^2-0$.
 Next, with $\tilde{m} = 2n-3$,
write  $\pi:\scr{X}\rightarrow\R^{\tilde{m}}$  for  that function which  assigns to
  $x = \matt{x_1'&x_2'&\cdots &x_n'}' \in \scr{X}$,
 the vector \footnotesize $$\matt{||x_2-x_1|| &(T_{(x_2-x_1)}(x_3-x_1))' &\cdots &(T_{(x_2-x_1)}(x_n-x_1))'}'$$ \normalsize
 in $\R^{\tilde{m}}$. Note that $\pi$ is well defined and smooth.
Clearly  $$||x_j-x_1||^2 =||T_{(x_2-x_1)}(x_j-x_1)||^2,\;\;\;j
\in\{3,4,\ldots,n\}$$ and
\begin{multline*}
||x_j-x_i||^2 =  ||T_{(x_2-x_1)}(x_j-x_1)-T_{(x_2-x_1)}(x_i-x_1)||^2,\\
i\in\{2,3,\ldots,n\},\;\;\;j\in\{i+1,,i+2,\ldots,n\}.
\end{multline*}
 Thus any entry in $\delta(x)$ is a polynomial function of entries in $\pi(x)$.
  Therefore there is a polynomial function
  $\bar{\delta}:\R^{\tilde{m}}\rightarrow\R^{\frac{n(n-1)}{2}}$ such that
  $\delta =\bar{\delta}\circ\pi$.
Since  same reasoning applies to the  error map  $\rho $,
 there must also
  be a polynomial function $\bar{\rho}:\R^{\tilde{m}}\rightarrow\R^{\tilde{m}}$ such that $\rho =\bar{\rho}\circ\pi$.

Note  that the derivative  of $\frac{1}{2}\rho$ at $y$, namely $\frac{1}{2}\frac{\partial \rho(x)}{\partial x}|_{x=y}$
 is the rigidity matrix of reduced   formation $\{\tilde{\mathbb{G}},y\}$. Since $\{\tilde{\mathbb{G}},y\}$ is minimally infinitesimally rigid,
 $\rank\frac{\partial \rho(x)}{\partial x}|_{x=y} = \tilde{m}$.  Meanwhile
$\frac{\partial \rho(x)}{\partial x}|_{x=y} =   \frac{\partial \bar{\rho}(q)}{\partial q}|_{q=\pi(y)}
\frac{\partial \pi(x)}{\partial x}|_{x=y} $.
Therefore \eq{\rank \frac{\partial \bar{\rho}(q)}{\partial q}|_{q=\pi(y)}\geq \tilde{m}.\label{l1}}
But
  $\frac{\partial \bar{\rho}(q)}{\partial q}|_{q=\pi(y)}$
 is an $\tilde{m}\times \tilde{m}$ matrix  so it must be nonsingular. Thus by the inverse function theorem, there is an open subset $\scr{W}\subset \R^{\tilde{m}}$  containing $\pi(y)$
  for which
  $\bar{\rho}$
  has a smooth
   inverse $\theta: \bar{\rho}(\scr{W}) \rightarrow \scr{W}$.
 Therefore
   $\theta(\bar{\rho}\circ\pi(x)) =\pi(x)$ for $\pi(x)\in\scr{W}$ or equivalently
  \eq{\theta(\rho(x)) =\pi(x),\;\;\;x\in\pi^{-1}(\scr{W}).\label{d1}}
 Note that the non-singularity of  $\frac{\partial \bar{\rho}(q)}{\partial q}$ at $q=\pi(y)$   implies that $\scr{W}$ can  be chosen so that \rep{d1} holds and at the same time, so that
 $\frac{\partial \bar{\rho}(q)}{\partial q}$ is nonsingular on $\scr{W}$. Let $\scr{W}$ be so defined.

 Set $\scr{A} = \pi^{-1}(\scr{W})$ and note that $y\in\scr{A}$ because $\pi(y)\in\scr{W}$.
 From \rep{d1}
 and the fact that $\delta = \bar{\delta}\circ\pi$  there follows
$$\delta(x) = \bar{\delta}\circ\theta\circ\rho(x),\;\;\;x\in\scr{A}.$$
Since $\rho(x) =\tilde{P}e(\bar{H}x),\;x\in\scr{A}$, \rep{func} holds with
$f=\bar{\delta}\circ\bar{\rho}^{-1}$.

The definition of $\pi$ implies that $\pi\circ\gamma $ for all transformations $\gamma$ in the restriction of $\mathfrak{G}$ to $\scr{X}$. This and the definition of $\scr{A}$ imply that $\scr{A}$ is $\mathfrak{G}$ - invariant.

 Non-singularity  of the matrix $\frac{\partial \bar{\rho}(q)}{\partial q}$ on $\scr{W}$
 implies non-singularity  of $\frac{\partial \bar{\rho}(q)}{\partial q}|_{q=\pi(x)}$ for $x\in\scr{A}$. Since
 the rigidity matrix of
  $\{\tilde{\mathbb{G}},x\}$ at $x\in\scr{A}$ can be written as $\frac{\partial \bar{\rho}(q)}{\partial q}|_{q=\pi(x)}
  \frac{\partial \pi(x)}{\partial x} $, to establish minimal infinitesimal rigidity of $\{\tilde{\mathbb{G}},x\}$  on $\scr{A}$, it is enough to show that for each  $x\in\scr{A}$
  \eq{\rank \frac{\partial \pi(x)}{\partial x}\geq \tilde{m}. \label{l23}}
  By direct calculation
  $$\frac{\partial \pi(x)}{\partial x} =  \matt{A_{1\times 4} & 0\cr\cr C_{(2n-4)\times 4} & B_{(2n-4)\times (2n-4)}}$$
  where $A= \frac{1}{||x1-x_2||}\matt{x_1'-x_2' & x_2'-x_1'}$, $ B = {\rm block \;diagonal\; }
  \{T_{x_2-x_1},\ldots,T_{x_2-x_1}\}$ and $C$ is some suitable defined matrix. Moreover $A$ is nonzero because $x\in\scr{X}$ and $B$ is nonsingular because $T_{x_2-x_1}$ is a rotation matrix. Thus the rows of $\frac{\partial \pi(x)}{\partial x}$ are linearly independent for $x\in\scr{A}$.
  It follows that \rep{l23} holds for all $x\in\scr{A}$ and thus $\{\tilde{\mathbb{G}},x\}$
  is minimally infinitesimally rigid for all such $x$
   $\qed $

\noindent{\bf Proof of Proposition \ref{quadlemma2}:}
 In view of Lemma \ref{pn2},  there exists a $\mathfrak{G}$ - invariant, open set $\scr{A}\subset\R^{2n}$ containing $y$ and a smooth function $f$  for which  \rep{func} holds.
  Let  $i,j,k,l$ be distinct integers in $\mathbf{n}$. In view of Lemma \ref{quadlemma}, to establish the correctness of statement  \rep{quad2} 
 it is enough to prove the existence of $\eta_{ijkl}$ for the case when $k =i$ and $j=l$.  But  the existence of  $\eta_{ijij} $ follows at once from \rep{func} and the definition of the squared distance function $\delta $. The minimal infinitesimally rigidity of $\{\tilde{\mathbb{G}},x\}$
 for $x\in\scr{A}$, follows  from Lemma \ref{pn2}.
  $\qed$

\subsection{Error System Definition}\label{esystemd}

A key step in the analysis of    the gradient law proposed in \cite{krickb}  is to show that along trajectories of
the overall system \rep{overall},
 the reduced error
vector $ \tilde{e}$  satisfies a self-contained
 differential equation of the form
$\dot{\epsilon} = g(\epsilon,\mu)$
where $g$ is a smooth function of just $\epsilon$ and $\mu$ and not $z$. As we will see, this can be
shown to be true  when $x(t)$  takes values in  the open subset $\scr{A}$ mentioned in the statement of Proposition \ref{quadlemma2}.
The precise technical result
is as follows.

\begin{theorem} Let $\{\mathbb{G},y\}$ be a target formation and let $\scr{A}$ be the open subset of $\R^{2n}$ mentioned in the statement of Proposition \ref{quadlemma2}. There exists a smooth function $g:\tilde{P}e(\bar{H}\scr{A})\times \R^{m}\rightarrow \R^{\tilde{m}}$
for which
\eq{g(\tilde{e},\mu)=-2\tilde{R}\tilde{R}'\tilde{e} -  2\tilde{R}\tilde{R}'F'(\tilde{e})\psi(\tilde{e}) + 2\tilde{R}S'\mu,\;\;x\in\scr{A}\label{voop}}
 where $\tilde{e}$ is the reduced error $\tilde{e} = \tilde{P}e(\bar{H}x)$ and $F$ is given by \rep{qoo3}. Moreover, if $x(t)$ is a solution to the overall system  \rep{overall}
for which $x(t)\in\scr{A}$ on some time interval $[t_0,t_1)$, then on the same time interval, the  reduced error vector  $\tilde{e} = \tilde{P}e(\bar{H}x(t))$ satisfies the self-contained differential equation
\eq{\dot{\epsilon} = g(\epsilon,\mu).\label{es}}
\label{china11}\end{theorem}
Although $\scr{A}$ and $g$ are defined for a specific target formation $\{\mathbb{G},y\}$, it is not difficult to see that both  are the same for all formations
which are in the same orbit as   $\{\mathbb{G},y\}$.
In the sequel we refer to \rep{es}  as the {\em error system} and we say that $\scr{A}$ is
   the {\em ambient space} on which it is valid.

\noindent{\bf Proof of Theorem \ref{china11}:}
The structures previously defined matrices
  $D(z) ={\rm diagonal}\; \{z_1,z_2,\ldots,z_m\}$,  $R(z) = D'(z)\bar{H}$, $S(z) = D'(z)\bar{J}$ $\tilde{R}(z) = \tilde{P}R(z)$  imply that  the entries of both $\tilde{R}(z)\tilde{R}'(z)$ and $\tilde{R}(z)S'(z)$
  are linear functions of the entries of the Gramian $\matt{z_1 & z_2 &\cdots z_m}'\matt{z_1 & z_2 &\cdots z_m}$.
    In view of \rep{pi} it is therefore clear that the entries of $\tilde{R}(z)\tilde{R}'(z)|_{z=\bar{H}x}$ and
  $\tilde{R}(z)S'(z)|_{z=\bar{H}x}$ can be written  as a  linear combination of  inner product terms of the form $(x_i-x_j)'(x_k-x_l)$ for $i,j,k,l\in\mathbf{n}$.  From this and Proposition \ref{quadlemma2} it is clear  that there exists an open subset $\scr{A}\subset\R^{2n}$  containing $y$ for which
each entry  in  $\tilde{R}(z)\tilde{R}'(z)|_{z=\bar{H}x}$ and
  $\tilde{R}(z)S'(z)|_{z=\bar{H}x}$ can be written as  a smooth function of $\tilde{P}e(\bar{H}x)$
    on $\tilde{P}e(\bar{H}\scr{A})$. The existence of a smooth function for which \rep{voop} holds follows at once. The second statement of the theorem is an immediate consequence is this and \rep{onion22}. $\qed $.


\subsection{Exponential Stability of the Unperturbed  Error System} \label{exun}

In this section we shall study the stability of the  error system  for the special case
  when $\mu =0 $. It is clear from \rep{voop} that in this case, the zero state $\epsilon = 0$
 is an equilibrium state of the unperturbed  error system.  The follow theorem states that this is in fact an
  {\em exponentially stable} equilibrium.

\begin{theorem}   The
equilibrium  state $\epsilon =0$ of the unperturbed error system
 $\dot{\epsilon} = g(\epsilon, 0)$  is  locally  exponentially stable. \label{mower}\end{theorem}

\noindent{\bf Proof of Theorem \ref{mower}:} First suppose that the target formation $\{\mathbb{G}, y\}$ is not minimally infinitesimally rigid, and that the reduced formation  $\{\tilde{\mathbb{G}},y\}$ is. To prove that $\epsilon=0$ is a locally exponentially stable equilibrium  it is enough to show that the linearization of
$\dot{\epsilon} = g(\epsilon,0)$ at $0$ is exponentially stable
 \cite{khalil2}. As noted in the proof of Theorem \ref{china11},  the matrix $\tilde{R}(z)\tilde{R}'(z)|_{z=\bar{H}x}$
  can be written as  a smooth function of $\tilde{P}e(\bar{H}x)$ on $\tilde{P}e(\bar{H}\scr{A})$
 Thus the function
$Q:\tilde{P}e(\bar{H}\scr{A})\rightarrow \R^{\tilde{m}\times \tilde{m}}$ for which $Q(\tilde{P}e(\bar{H}x)) = \tilde{R}(\bar{H}x)\tilde{R}'(\bar{H}x)$ is well defined, smooth, and  positive semi-definite on $\tilde{P}e(\bar{H}\scr{A}) $.
We claim that $Q(0)$ is nonsingular and thus positive definite.  To understand why this is so,
recall that for any vector $x\in\R^{2n}$,
$\tilde{R}(z)|_{z=\bar{H}x} $ is the rigidity matrix
of the   formation $\{\tilde{\mathbb{G}},x\}$. In addition, $y\in\scr{A}$ so by Proposition \ref{quadlemma2},
 the formation  $\{\tilde{\mathbb{G}},y\}$ is minimally infinitesimally rigid.  Therefore
$\rank \tilde{R}(\bar{H}y) = \tilde{m} $. Hence  $\tilde{R}(\bar{H}y)\tilde{R}'(\bar{H}y)$ is nonsingular. Moreover
 $Q(0) = \tilde{R}(\bar{H}y)\tilde{R}'(\bar{H}y)$ since  $\tilde{e}(\bar{H}y) = 0$. Therefore $Q(0)$ is nonsingular as claimed.

From  \rep{voop}, the definition of $Q$, and the definition of  $F$ in \rep{qoo3}, it is clear that for $\epsilon\in
\tilde{P}e(\bar{H}\scr{A})$
$$\frac{\partial g(\epsilon,0)}{\partial \epsilon}=
 -2Q(I+F'F +F'\psi ) -2\left (\frac{\partial Q}{\partial \epsilon}\right )(\epsilon +F'\psi).$$
Therefore \eq{\left. \frac{\partial g(\epsilon,0)}{\partial \epsilon}\right |_{\epsilon = 0}=  -2Q(0)(I + F'(0)F(0)).\label{apple}}
But $-2Q(0)(I + F'(0)F(0))$ is similar to $-2TQ(0)T'$ where $T$ is any nonsingular matrix such that $T'T = I + F'(0)F(0)$. Since  $-2TQ(0)T'$ is negative definite, it is a stability matrix.
 Hence $-2Q(0)(I + F'(0)F(  0))$ is a stability matrix.
 Therefore the linearization of  $\dot{\epsilon} = g(\epsilon,0)$ at $\epsilon =0$
is exponentially stable. Therefore
 $\epsilon =0$ is an exponentially stable equilibrium of the error system $\dot{\epsilon} = g(\epsilon,0)$.

 Now suppose that  $\{\mathbb{G},y\}$  itself is minimally infinitesimally rigid. In this case
  the same argument just used applies except that in this case, the right side of \rep{apple}
    is just $-2Q(0)$.  $\qed $

\subsection{Exponential Stability of the Perturbed  Error System}\label{expert}

As is well known, a critically important property of exponential stability is  {\em robustness}.
We now explain exactly what this means for the  error system under consideration. First suppose that the target formation $\{\mathbb{G}, y\}$ is not minimally infinitesimally rigid, and that the reduced formation  $\{\tilde{\mathbb{G}},y\}$ is.

\begin{enumerate}\item As was just  shown in the proof of Theorem \ref{mower}, \rep{apple} holds
and $-2Q(0)(I + F'(0)F(0))$ is a stability matrix.

\item Clearly   $-2Q(0)(I + F'(0)F(0))$ is nonsingular. Therefore by the implicit function theorem, there exists
  an open neighborhood $\scr{M}_0\subset\R^{m}$  centered at $0$
  and a vector  $\epsilon_{\mu} $ which is a smooth function of $\mu$  on $\scr{M}_0$ such that $\epsilon_0 = 0$ and
   $g(\epsilon_{\mu},\mu) = 0$
   for $\mu\in\scr{M}_0$.

\item The Jacobian matrix
$$J(\mu)=\frac{\partial g(\epsilon, \mu)}{\partial \epsilon} |_{\epsilon = \epsilon_{\mu}}$$
in continuous on $\scr{M}_0$ and equals the  stability matrix $-2Q(0)(I + F'(0)F(0))$
at $\mu = 0$.
For any $\mu$ in any sufficiently small open  neighborhood
$\scr{M}\subset \scr{M}_0$ about $0$,   $J(\mu )$  is a stability matrix.

\item Stability of $J(\mu)$ is equivalent to exponential stability of the equilibrium  state
$\epsilon_{\mu}$
of the  system $\dot{\epsilon} = g(\epsilon , \mu)$.

   \end{enumerate}
   Of course the same arguments apply, with minor modification, to the case when  $\{\mathbb{G}, y\}$ itself is  minimally infinitesimally rigid.
We summarize:

\begin{corollary} On any sufficiently small  open neighborhood $\scr{M}\subset\R^m$ about $\mu = 0$, there is  a smooth function
$\mu\longmapsto \epsilon_{\mu} $  such that $\epsilon_0 = 0$ and for each $\mu\in\scr{M}$, $\epsilon_{\mu} $
is an exponentially stable equilibrium  state of the  error system
 $\dot{\epsilon} = g(\epsilon,\mu).$
 \label{frunch}\end{corollary}

Prompted by \rep{pizzas}, we define the {\em equilibrium output} of the error system  $\dot{\epsilon} = g(\epsilon,\mu)$ to be
$\mathfrak{e}_{\mu} =  \tilde{P}'\epsilon_{\mu} +\widehat{P}'\psi(\epsilon_{\mu})$. As noted just below the statement of Proposition \ref{quadlemma2}, $\psi$ is a smooth function and $\psi(0) = 0$. Thus, like
the error system's equilibrium state, $\mathfrak{e}_{\mu}$ is a smooth function of $\mu$ and $\mathfrak{e}_{0} = 0$.

\subsection{Exponential Convergence}\label{expos}

At this point we have shown that for $\mu\in\scr{M}$, the equilibrium $\epsilon_{\mu}$ of the error system is locally exponentially stable. We have also shown that along any trajectory of the overall system for which $x(t)\in\scr{A}$,  the reduced error  $\tilde{e}$ satisfies the error  equation \rep{es} and the overall error satisfies \rep{pizzas}.
 It remains to be shown that if   $x$ starts out at a value in some suitably defined open subset of $\scr{A}$ for which $\tilde{P}e(\bar{H}x)$ is within the domain of attraction of the error system's equilibrium $\epsilon_{\mu}$, then $x$ will remain  within the subset $\scr{A}$  for all time and consequently
 $\tilde{e}$ and $e$  and will  converge exponentially fast to $\epsilon_{\mu}$  and   $\mathfrak{e}_{\mu} = \tilde{P}'\epsilon_{\mu} +\widehat{P}'\psi(\epsilon_{\mu})$ respectively. This is the subject of Theorem \ref{opp}  below.

Before stating the theorem we want to emphasize that just because the reduced error might start out at  a value  $\tilde{P}e(\bar{H}x(0))$  which is close to  $\epsilon_{\mu}$  or even equal to $\epsilon_{\mu}$, there is no guarantee that $x(0)$ will be in $\scr{A}$. In fact the {\em only} situation
when  $\tilde{P}e(\bar{H}x(0)) = 0$  would imply $x\in\scr{A}$ is when the target formation is globally rigid \cite{hendrickson}. The complexity of this entire problem can be traced to this point. The problem being addressed here {\em cannot} be treated  as a standard local stability problem in error space.

\begin{theorem}
  Let $\{\mathbb{G},y\}$ be a target formation and let   $\scr{A}$   be the opened set referred to in the statement of   Proposition \ref{quadlemma2}.
 For each value of  $\mu$ in any  sufficiently small open neighborhood $\scr{M}$ in $\R^m $ about $\mu =0$,  and each initial state $x(0)\in\scr{A}$ for which the  error
 $e(\bar{H}x(0))$ is sufficiently close to the equilibrium output $\mathfrak{e}_{\mu}$ of the error system  $\dot{\epsilon} = g(\epsilon, \mu)$, the following statements are true:
    \begin{enumerate}
    \item The trajectory of the overall system starting at $x(0)$ exists for all time and lies in $\scr{A}$. \label{cm1}
     \item The error $e=e(\bar{H}x(t))$  converges exponentially fast to $\mathfrak{e}_{\mu}$.
     \label{cm3}\end{enumerate}
  \label{opp}\end{theorem}
  To prove this theorem, we will need the following lemmas.

\begin{lemma} Let $\{\mathbb{G},y\}$ be a target formation and let   $\scr{A}$   be the opened set referred to in the statements of Lemma \ref{pn2}  and Proposition \ref{quadlemma2}.
There exists an open ball  $\scr{B}_o\subset \R^{\tilde{m}}$ centered at $0$ and an
   open set $\scr{C}\subset \R^{2n}$ such that  $\scr{C}$ and the closure of $\scr{A}$ are disjoint
   and
    \eq{\rho^{-1}(\scr{B}_o) \subset \scr{A} \cup\scr{C}.\label{pan}}
  \label{xmas}\end{lemma}

\noindent{\bf Proof of Lemma \ref{xmas}}:
Let  $\{\mathbb{G},y\},\{\mathbb{G},y^2\},\ldots, \{\mathbb{G},y^{n_o}\}$ be representative formations within
  the $n_o$ disjoint orbits whose union is the set of all formations equivalent to  $\{\mathbb{G},y\}$. By assumption, each of these formations is infinitesimally rigid. Thus by the same reasoning used to establish the existence of $\scr{W}$ and $\theta$  in the proof of Lemma
   \ref{pn2}, one can conclude that for each $i\in\{2,3,\ldots,n_o\}$, there is an open subset
   $\scr{W}_i\subset\R^{\tilde{m}}$ containing $\pi(y^i)$ and an inverse function $\theta_i:\bar{\rho}(\scr{W}_i) \rightarrow \scr{W}_i$.  Since the $\pi(y^i)$ and $\pi(y)$
   are distinct points , the $\scr{W}_i$ can be assumed to have been chosen small enough so that
   all are disjoint with  the closure of $\scr{W}$. Thus the set $\scr{S} = \cup_{i=2}^{n_o}\scr{W}_i$ and the closure of $\scr{W}$ are disjoint.


  Note that the restriction of $\bar{\rho}$ to $\scr{W}$ is $\theta^{-1} $ which in turn is a homeomorphism.  Thus the restriction of  $\bar{\rho}$ to $\scr{W}$ is  an open function which implies that $\bar{\rho}(\scr{W})$ is an opened set. By similar reasoning, each
  $\bar{\rho}(\scr{W}_i)$ is also an opened set as is the
  intersection $\cap_{i\in \mathbf{n_o}}\bar{\rho}(\scr{W}_i)$ where $\scr{W}_1 \dfb \scr{W}$. Moreover
  $0$ is in this intersection because $\pi(y)\in \scr{W}_1$, $\pi(y^i) \in \scr{W}_i,\;i\in\{2,3,\ldots,n_o\}$ and $\bar{\rho}$ maps each of these points into $0$.

Let $\scr{T}$ be the complement of $\scr{W}\cup\scr{S} $ in $\R^{\tilde{m}}$.
Note that $\bar{\rho}(\scr{T})$ cannot
contain the origin because  the only points in the domain of $\bar{\rho}$ which map into $0$  are in $\scr{W}\cup\scr{S} $. This implies that the set
$$\scr{B}_0= \bigcap_{i\in \mathbf{n_o}}\bar{\rho}(\scr{W}_i ) -\bar{\rho}(\scr{T} )\cap\bigcap_{i\in \mathbf{n_0}}\bar{\rho}(\scr{W}_i )$$
contains $0$. We claim that $\scr{B}_0$ is opened. Since $\cap_{i\in\mathbf{n_o}}\bar{\rho}(\scr{W}_i)$ is opened,
 to establish this, it is enough to show that $\scr{T}\cap\cap_{i\in\mathbf{n_o}}\bar{\rho}(\scr{W}_i)$ is closed. This in turn will be true if  $\bar{\rho}(\scr{T} )$
is closed.  But this is so because $\scr{T}$ is closed and because $\bar{\rho}$ is a weakly coercive, polynomial function mapping one finite dimensional vector space into another \cite{coercive}.

We claim that
\eq{\bar{\rho}^{-1}(\scr{B}_o)\subset \scr{W}\cup\scr{S}.\label{vac}}
To show that this is so, pick $q\in\bar{\rho}^{-1}(\scr{B}_o)$ in which case $\bar{\rho}(q)\in\scr{B}_o$. But $q\subset \R^{\tilde{m}}$ and $\R^{\tilde{m}} =
\scr{W}\cup\scr{S}\cup\scr{T}$ so $q\in\scr{W}\cup\scr{S}\cup\scr{T}$.
But $q$ cannot be in $\scr{T}$ because $\bar{\rho}(q)\in\scr{B}_o$ and $\scr{B}_o$ and $\bar{\rho}(\scr{T})$ are disjoint. Thus $\bar{\rho}(q)\in \scr{W}\cup\scr{S} $ so  $q\in \bar{\rho}^{-1}(\scr{W}\cup\scr{S}) $. Therefore \rep{vac} is true.

We claim that \rep{pan} holds with $\scr{C}= \pi^{-1}(\scr{S})$ and that with this choice, $\scr{C}$ and the closure of $\scr{A}$ are disjoint. To establish \rep{pan}, pick $x\in\rho^{-1}(\scr{B}_o)$;
then $\pi(x)\in\bar{\rho}^{-1}(\scr{B}_o)$   because $\rho = \bar{\rho}\circ\pi$. In view of \rep{vac}, $\pi(x)\in\scr{W}\cup\scr{S}$. But $\scr{A} = \pi^{-1}(\scr{W})$ and $\scr{C}= \pi^{-1}(\scr{S})$ so $x\in \scr{A}\cup\scr{C}$. Therefore  \rep{pan} is true.

To complete the proof we need to show that $\scr{C}$ and the closure of $\scr{A}$ are disjoint.
Note that because $\pi$ is continuous,  $\bar{\scr{A}}\subset \pi^{-1}(\bar{\scr{W}})$ where $\bar{\scr{A}}$ and $\bar{\scr{W}}$
are the closures of $\scr{A}$ and $\scr{W}$ respectively.
Then  $\bar{\scr{A}}\cap\scr{C}\subset \pi^{-1}(\bar{\scr{W}})\cap\pi^{-1}(\scr{S})\subset
\pi^{-1}(\bar{\scr{W}}\cap\scr{S})$.  But  $\bar{\scr{W}} $  and $\scr{S}$ are disjoint so
$\bar{\scr{A}} $  and $\scr{C}$  must be disjoint as well. $\qed$

  \begin{lemma} Let $\{\mathbb{G},y\}$ be a target formation and let $\scr{A}$ be the opened set referred to in the statement of Lemma \ref{pn2}  and Proposition \ref{quadlemma2}. For any sufficiently small open ball $\scr{B}\subset \R^{\tilde{m}}$ centered at $0$,
  \eq{\scr{B} \subset \tilde{P}e(\bar{H}\scr{A})\label{blast}}
  and
  \eq{{\rm closure}(\scr{A}_{\scr{B}})\subset \scr{A} \label{blast2}}
where
\eq{\scr{A}_{\scr{B}}  =\{x:\tilde{P}e(\bar{H}x)\in\scr{B},\;x\in\scr{A}\}.\label{blaster}}
\label{xmasday}\end{lemma}

\noindent{\bf Proof of Lemma \ref{xmasday}}: Since $\rho=\bar{\rho}\circ \pi$,
$\rho(\scr{A}) = \bar{\rho}(\pi(\scr{A}))$.  Moreover $\scr{A} = \pi^{-1}(\scr{W})$; but $\pi$ is surjective  so $\pi (\scr{A}) = \scr{W}$. As noted in the proof
 of Lemma \ref{xmas}, the restriction of $\bar{\rho} $ to $\scr{W}$ is an open map.  Since
 $\rho(\scr{A}) = \bar{\rho}(\scr{W})$ and $\scr{W}$ is opened, $\rho(\scr{A})$ must be opened as well. Furthermore, $y\in\scr{A}$ and $\rho(y) = 0$, so $0\in \rho(\scr{A})$. But by definition,
  $\tilde{P}e(\bar{H}\scr{A}) = \rho(\scr{A})$ so  $\tilde{P}e(\bar{H}\scr{A})$ is also opened and contains $0$.   Thus \rep{blast} will hold provided $\scr{B}$ is sufficiently small.

Let $\scr{B}_0$ be as in the statement of Lemma \ref{xmas}. Let $\scr{B}$ be any opened ball in $\R^{\tilde{m}}$ centered at $0$ which is small enough so that its closure $\bar{\scr{B}}$ is contained in $\scr{B}_0$. Let $x_1,x_2,\ldots, x_i ,\ldots $ be a convergent
 sequence in $\scr{A}_{\scr{B}}$
  with limit $x^*$.  To establish \rep{blast2}, it is enough to
 show that $x^*\in\scr{A}$.  By definition, $\scr{A}_{\scr{B}}=\scr{A}\cap\rho^{-1}(\scr{B}) $
 so $x_i\in \scr{A}\cap\rho^{-1}(\scr{B}),\;i\geq 1$. Therefore
  $x^*\in\bar{\scr{A}}$ which is the closure of $\scr{A}$.  Since $x_i\in\rho^{-1}(\scr{B})$, $\rho(x_i)\in \scr{B}$.  Therefore $\rho(x^*)\in\bar{\scr{B}}$ so  $\rho(x^*)\in\scr{B}_o$.
  Hence
  $x^*\in\rho^{-1}(\scr{B}_0)$. Thus by Lemma \ref{xmas},
 $x^*\in \scr{A}\cup\scr{C}$.  But $x^*\in\bar{\scr{A}}$ and $\bar{\scr{A}}$ and $\scr{C}$ are disjoint so $x^*\in\scr{A}$.  Therefore \rep{blast2} is true.
$\qed $

   \noindent{\bf Proof of Theorem \ref{opp}:} Let $\scr{B}$ be an open ball centered at $0$ which satisfies  \rep{blast} and \rep{blast2}.
  Corollary \ref{frunch} guarantees that  for any $\mu$ in any sufficiently small open  neighborhood
$\scr{M}$ about $0$, the  error system  has an exponentially stable equilibrium $\epsilon_{\mu}$.
Suppose $\scr{M}$ is any such neighborhood, which is also  small enough so that
for each $\mu\in\scr{M}$,  $\epsilon_{\mu}\in \scr{B}$.
Since  for each $\mu \in \scr{M}$, the  error system has $\epsilon_{\mu}$ as an exponentially stable equilibrium, for each such $\mu$ there must be a sufficiently small positive radius $r_{\mu}$ for which any trajectory of the error system  starting in   $\{\epsilon:||\epsilon-\epsilon_{\mu}|| <r_{\mu}\}$,
 lies wholly  within $\scr{B}$ and converges to $\epsilon_{\mu}$ exponentially fast.

  Now fix $\mu\in\scr{M}$.
   We claim that for any point $\epsilon\in\R^{\tilde{m}}$ such that
  $||\epsilon - \epsilon_{\mu}||<r_{\mu}$, there is at least one vector $q\in\scr{A}$ for which $\tilde{P}e(\bar{H}q) = \epsilon $. To establish this claim, note first that $\epsilon\in\scr{B}$.
  In view of \rep{blast}, there must be a vector $p\in\scr{A}$ such that
  $\tilde{P}e(\bar{H}p) = \epsilon $.  
 Thus $q=p$ has the required property.

 Now let
  $q$ be any state in $\scr{A}$  such that
   $e(\bar{H}q) $ is close enough to $\mathfrak{e}_{\mu}$ so that $||\tilde{P}e(\bar{H}q)-\epsilon_{\mu}||<r_{\mu}$.
  Then $\tilde{P}e(\bar{H}q)\in\scr{B}$ so  $q\in\scr{A}_{\scr{B}}$. 
Let  $x(t)$ be the solution to the overall system starting in state $q$. Then $x(0)\in\scr{A}_{\scr{B}}$.
Let $[0,T)$ denote the maximal interval of existence for this solution and let   $T^*$ denote the largest
time  in this interval such that $x(t)\in\scr{A}_{\scr{B}}$  for $t\in[0,T^*)$. Suppose $T^* <T$ in which  case $x(t)$ is well defined on the closed interval $[0,T^*]$. Moreover, since
$x(t)$ is continuous and in $\scr{A}_{\scr{B}}$ on $[0,T^*)$ it must be true that
 $x(t)\in {\rm closure }(\scr{A}_{\scr{B}}),\;t\in[0,T^*]$.  Therefore
  $x(t) \in\scr{A},\;\;t\in[0,T^*]$ because of \rep{blast2}.
 In view of  Theorem \ref{china11},
  $\tilde{P}e(\bar{H}x(t)) = \epsilon(t),\;t\in[0,T^*] $ where $\epsilon(t)$  is the solution to the error system
 starting at $\epsilon(0) = \tilde{e}(\bar{H}q) $. But  $||\epsilon(0) -\epsilon_{\mu}||<r_{\mu}$,  so
  for $t\in[0,T^*]$,
 $\epsilon(t) \in \scr{B}$. Therefore  $\tilde{P}e(\bar{H}x(t))\in\scr{B}$ for $t\in[0,T^*]$.
  It follows that    $x(t)\in\scr{A}_{\scr{B}},\;t\in[0,T^*]$. This
contradicts the hypothesis that  $T^*$
is the largest time such that  $x(T^*)\in\scr{A}_{\scr{B}}$ for $t<T^*$.
Therefore $T^*=T$. 
Clearly $\tilde{P}e(\bar{H}x(t)) = \epsilon(t) $ and $x(t)\in\scr{A}_{\scr{B}}$ for all $t\in[0,T) $.  

 The definition of  $r_{\mu}$ and the assumption that $||\epsilon(0) - \epsilon_{\mu}||<r_{\mu}$  imply that  $\epsilon(t)$ must converge to $\epsilon_{\mu}$ exponentially fast. Thus there must be a positive constant $c$ such that $||\epsilon(t)||\leq c,\;t\geq 0$.
 Therefore $||\tilde{P}e(\bar{H}x(t))||\leq c,\;t\in[0,T)$. In view of \rep{pizzas},
  $||e(\bar{H}x(t))||\leq \bar{c}$  where $\bar{c} = ||\tilde{P}'||c +||\widehat{P}'||\sup_{||\zeta||\leq c}||\psi(\zeta)||$.
   Clearly $||e_k(z(t))||  \leq \bar{c},\;\;k\in\mathbf{m}$.
  Therefore   $||z_k(t)|| \leq  \sqrt{\bar{c} +d^2_k},\;\;\;k\in\mathbf{m}$
  because of \rep{e}. Therefore $||z(t)||$ must be bounded on $[0,T)$ by a constant $C$ depending only on $\bar{c}$ and the $d_k$.  Since $\dot{x} =  -R'(z)e(z)+S'(z)\mu$ and $R(\cdot)$ and $S(\cdot)$ are continuous, it must be true that $||\dot{x}||$
  is bounded on $[0,T)$  by a finite constant.
  This implies that $||x(t)|| \leq c_1 +c_2T,\;t\in[0,T)$  where $c_1$ and $c_2$ are constants.
   From this it follows by a standard argument that $T=\infty$. Therefore statement \ref{cm1} of the theorem is true. 
   Statement \ref{cm3} is a consequence of
   \rep{pizzas}  and the fact that  $\tilde{P}e(\bar{H}x(t))  = \epsilon(t),\;t\geq 0$. $\qed$

The proof of Theorem \ref{opp} makes it clear that if  the overall
 system starts out  in a state $x(0)\in\scr{A}$ for which the  error $e(\bar{H}x(0))$  is sufficiently close to the error system  equilibrium output $\mathfrak{e}_{\mu}$, then the trajectory of the overall system lies wholly with $\scr{A}_{\scr{B}}$ for all time. Repeated use of this fact will be made throughout the remainder of this paper.

\section{Square Subsystem}\label{redu}

In this section we derive
 a special $2\times 2$ ``square'' sub-system whose behavior along  trajectories of the overall system will enable us to easily predict the behavior of $x(t)$.
Suppose that  during some time  period $[t_0,t_1)$, the state $x(t)$  of the overall system is ``close''  to a state $y$ for which $\{\mathbb{G},y\}$ is a target formation. Since sufficient closeness of $x(t)$ and $y$   would mean that  the formation in $\{\mathbb{G},x(t)\}$ is infinitesimally rigid,
  it is natural to expect that if $x(t)$ and $y$ are close enough during  the  period $[t_0,t_1)$, then over this period the behavior of all of the $z_i$ will depend on only a few of the $z_i$. As we will soon see, this is indeed the case. To explain why this is so,
 we will make use of the  fact that the $z$ system in  \rep{z2} can  also  be written as
\eq{\matt{\dot{z}_1 & \dot{z}_2 & \cdots & \dot{z}_m} =\matt{z_1 & z_2 & \cdots & z_m}M(e(z),\mu)\label{pup}}
 where $M(e,\mu)$ is a $m\times m$ matrix  depending linearly on the pair $(e,\mu)$. This is a direct consequence of the definition of the $z_i$ in \rep{pi} and the fact that the $x_i$ satisfy
\rep{ov}.
   We can now state the following proposition.

\begin{proposition}
Let $\{\mathbb{G},y\}$ be a target formation.
 There are integers  $p,q\in\mathbf{m}$  depending on $y$ for which $z_p(y)$ and $z_q(y)$ are linearly independent. Moreover, for any
    ball $\scr{B}$ about  zero which satisfies \rep{blast} and \rep{blast2},   the matrix $Z(z) = \matt{z_p &z_q}$ is nonsingular on $\bar{H}\scr{A}_{\scr{B}}$ and  there is    a smooth
 matrix - valued function $Q:e(\bar{H}\scr{A}_{\scr{B}})\rightarrow \R^{2\times m}$ for which
\eq{\matt{z_1 &z_2 &\ldots &z_m} =Z(z) Q(e(z)),\;\;z\in\bar{H}\scr{A}_{\scr{B}}.\label{sb1}}
If $x(t)$ is a solution to the overall system  \rep{overall}
for which $x(t)\in\scr{A}_{\scr{B}}$ on some time interval $[t_0,t_1)$, then on the same time interval,
$Z(\bar{H}x(t))$ is nonsingular and satisfies
 \eq{\dot{Z} = ZA(e(z),\mu)\label{sq1}}
where $A(e,\mu) = Q(e)M(e,\mu)L$ and $L$ is the $m\times 2 $ matrix  whose columns are the $p$th and $q$th unit vectors in $\R^m$. Moreover \eq{Q(\mathfrak{e}_{\mu})M(\mathfrak{e}_{\mu},\mu) = A(\mathfrak{e}_{\mu},\mu)Q(\mathfrak{e}_{\mu})\label{pupps}}
where $\mathfrak{e}_{\mu}$ is the equilibrium output of the error system.
\label{resubmit}\end{proposition}
The proposition clearly implies that  on the time interval $[t_0,t_1)$, the behavior of the entire vector $z$ is determined by the behavior of the
 {\em square subsystem}  defined by \rep{sb1} and \rep{sq1}.
The proof of Proposition \ref{resubmit} depends on  Lemma \ref{china2} which we state below.

In the sequel we write  $p_1\wedge p_2$ for  the {\em wedge product}
$p_1\wedge p_2 = \det \matt{p_1 &p_2}$ of any two vectors $p_1,p_2\in\R^2$. The wedge product is a bilinear map.

\begin{lemma} Let $\{\mathbb{G},x\}$ be an infinitesimally rigid formation in $\R^2$  with multi-point
$x=\matt{x'_1 & x'_2 & \cdots &x'_n}'$. Then for at least one pair of edges $(i,k)$ and $(j,k)$ in $\mathbb{G}$ which share a common vertex $k$,   $(x_i-x_k)\wedge (x_j-x_k) \neq 0$. \label{china2} \end{lemma}
A proof of this lemma is given in the appendix.

\noindent{\bf Proof of Proposition \ref{resubmit}:} Since $\{\mathbb{G},y\}$ is a target formation, it is infinitesimally rigid.  In view of  Lemma \ref{china2} and the definition of $z_i$ in \rep{pi}, there exist integers $p,q\in\mathbf{m}$
for which $z_p\wedge z_q \neq 0$ at $x=y$. By a simple computation
\eq{(z_p\wedge z_q)^2 = ||z_p||^2||z_q||^2 - (z_p'z_q)^2\label{www}} for all $x$. Suppose that $x_i,x_j,x_k,x_l$ are the sub-vectors of $x$ for which $z_p=x_i-x_j$ and $z_q=x_k-x_l$. As a consequence of    \rep{www} and Proposition \ref{quadlemma2}
there
is a smooth function $h:e(\bar{H}\scr{A})\rightarrow\R$ such that $(z_p\wedge z_q)^2 = h(e(\bar{H}x)),\;x\in\scr{A}$; moreover  $h(e(\bar{H}y))\neq 0$ because $z_p(y)\wedge z_q(y) \neq 0$.  It follows that  $z_p\wedge z_q \neq 0,\;x\in\scr{A}_{\scr{B}}$. Thus $z_{p}\wedge z_q\neq 0$ for all $z\in\bar{H}\scr{A}_{\scr{B}}$  so the matrix
   $Z(z)=\matt{z_p &z_q}$ is nonsingular for all such $z$.

To proceed, note  that for all  $z\in \bar{H}\scr{A}_{\scr{B}}$,  the matrix
  $$P(z) =   \matt{z_p &z_q}^{-1}\matt{z_1 &z_2 &\cdots &z_m}$$ satisfies
  $$\matt{z_1 &z_2 &\cdots &z_m} = \matt{z_p &z_q} P(z).$$
  We claim that   for   $z\in \bar{H}\scr{A}_{\scr{B}}$,  $P(z)$
 depends only on $e(z)$; that is $P(z) = Q(e(z))$  for some matrix $Q$ which is a smooth function of $e$.
To establish this claim, note  first that $P(z)$ can be written at
$$P(z) =\matt{z_p &z_q}^{-1} (K\matt{z_p &z_q})(K\matt{z_p &z_q})^{-1}
\cdot \matt{z_1  &z_2 &\ldots &z_m}$$
where $K=\matt{0 & -1\cr 1 &0}$.
 By Cramer's rule
  $$(KZ)^{-1}\matt{z_1 &z_2 &\dots & z_m}=     \frac{1}{z_p\wedge  z_q}\matt{z_1\wedge (Kz_q) &z_2\wedge (Kz_q) &\cdots &z_m\wedge (Kz_q)\cr
                (Kz_p)\wedge z_1 &(K z_p)\wedge z_2 &\cdots &(Kz_p)\wedge z_m}$$
                and
$$ Z^{-1}(KZ)= \frac{1}{z_p\wedge z_q}\matt{(Kz_p)\wedge z_q &(Kz_q)\wedge z_q\cr  z_p\wedge (Kz_p) & z_p \wedge (Kz_q)}.$$
But for all $i,j\in\mathbf{m}$,  $z_i\wedge (K z_j) = z_i'z_j$ and $(Kz_i)\wedge z_j = -z_i'z_j$. From this and \rep{www} it follows that
$$P(z) = \frac{1}{(||z_p||^2||z_q||^2 - (z_p'z_q)^2)}\matt{z_p'z_q &z_q'z_q\cr -z_p'z_p &-z_p'z_q}\matt{-z_1'z_q &-z_2'z_q &\cdots & -z_m'z_q\cr z_p'z_1 &z_p'z_2 &\cdots & z_p'z_m}$$
Since $||z_p||^2||z_q||^2 - (z_p'z_q)^2\neq 0, z\in\bar{H}\scr{A}_{\scr{B}}$,
  $P(z)$  is a smooth function of
terms of the form $z_r'z_s,\;r,s\in\mathbf{m}$.

In view of \rep{pi} it is therefore clear that  on $\scr{A}_{\scr{B}}$, $P(\bar{H}x)$ is a smooth function of
     inner products of the form $(x_i-x_j)'(x_k-x_l)$ for $i,j,k,l\in\mathbf{n}$. But $\scr{A}_{\scr{B}}\subset\scr{A}$.
    From this and Proposition \ref{quadlemma2} it therefore follows that on $\scr{A}_{\scr{B}}$,
     $P(\bar{H}x) = Q(e(\bar{H}x))$ where on $e(\bar{H}\scr{A}_{\scr{B}})$,
       $Q(e)$ is a smooth function of $e$.   Thus the claim is established and  \rep{sb1} is true.
Moreover, since  $x(t)\in\scr{A}_{\scr{B}}$ for $t\in[t_0,t_1)$, $z\in \bar{H}\scr{A}_{\scr{B}}$ on the same time interval. Therefore $Z$ is nonsingular on $[t_0,t_1$).

 The definition of $L$ implies that \eq{Z = \matt{z_1 &z_2 &\ldots &z_m}L .\label{fall}}
 Thus from \rep{pup},
  \eq{\dot{Z} = \matt{z_1 &z_2 &\ldots &z_m}M(e,\mu)L.}
This and \rep{sb1} imply that \rep{sq1} is true.

To prove \rep{pupps},  let $x(t)$ be a solution to the overall system in $\scr{A}$, along which $e(\bar{H}x) = \mathfrak{e}_{\mu}$. In view of Theorem \ref{opp}, such a solution exists. Moreover  $x(t)\in \scr{A}_{\scr{B}},\;t\geq 0$. Clearly $z(t)\in \bar{H}\scr{A}_{\scr{B}},\;t\geq 0$ so
\eq{\matt{z_1 &z_2 &\ldots &z_m} = Z(z)Q(\mathfrak{e}_{\mu}),\;t\geq 0\label{tops}}
because of \rep{sb1}.
We claim that
\eq{\ker Q(\mathfrak{e}_{\mu}) \subset  \ker Q(\epsilon_{\mu})M(\mathfrak{e}_{\mu},\mu).\label{daily}}
To prove this claim, let $p$ be any vector such that $Q(\mathfrak{e}_{\mu})p=0$. Then $ZQ(\mathfrak{e}_{\mu})p=0,\;t\geq 0$  \ so
$$\matt{z_1 &z_2 &\ldots &z_m}p=0,\;t\geq 0$$
  because of \rep{tops}. Therefore
  $$\matt{\dot{z}_1 & \dot{z}_2 & \cdots &\dot{z}_m}p=0,\;t\geq 0$$ so
   $$\matt{z_1 &z_2 &\ldots &z_m}M(\mathfrak{e}_{\mu},\mu)p=0,\;t\geq 0$$
   because of \rep{pup}.  Hence by \rep{tops},
   $ZQ(\mathfrak{e}_{\mu})M(\mathfrak{e}_{\mu},\mu)p =0,\;t\geq 0$. But $Z$ is nonsingular for $t\geq 0$, so
  $Q(\mathfrak{e}_{\mu})M(\mathfrak{e}_{\mu},\mu)p =0$.  Since $p$ is arbitrary, \rep{daily} is true.

In view of \rep{daily}, there must be a matrix $B$ such that $Q(\mathfrak{e}_{\mu})M(\mathfrak{e}_{\mu},\mu) = BQ(\mathfrak{e}_{\mu})$
But from  \rep{fall} and \rep{sb1} we see that $Z=ZQL$.  Since $Z$ is nonsingular, $QL = I_{2\times 2}$. Thus
$Q(\mathfrak{e}_{\mu})M(\mathfrak{e}_{\mu},\mu)L = BQ(\mathfrak{e}_{\mu})L = B$, so $B=A(\mathfrak{e}_{\mu},\mu)$. $\qed $

\section{Analysis of the Overall System}\label{anals}

In view of Theorem  \ref{opp}, we now know that for any mismatch error $\mu$ with small norm and any initial state $x(0)\in\scr{A}$ for which
 $e(\bar{H}x(0))$ is close to the error system equilibrium output $\mathfrak{e}_{\mu}$, the error signal $e(\bar{H}x(t))$ must converge exponentially fast to  $\mathfrak{e}_{\mu}$ and $\dot{x}(t)$ must be bounded on $[0,\infty)$.  But what about $x(t)$ itself?  The aim of the remainder of this paper is to answer this question.  We will address the question in two steps.  First in Section \ref{step1} we will consider the situation when
$e(\bar{H}x(t))$ has already converged  $\mathfrak{e}_{\mu}$.   Then in Section \ref{NEA} we will elaborate on the case  when $e(\bar{H}x(t))$ starts out close to $\mathfrak{e}_{\mu}$.

\subsection{Equilibrium Analysis}\label{step1}
 The aim of this section is to  determine the behavior of the  formation $\{\mathbb{G},x(t)\}$ over  time  for $x(t)\in\scr{A}$ and for mismatch errors from a suitably defined  ``generic'' set, assuming  that  for each such value of $\mu$,
  the  error $e(\bar{H}x(t))$  is constant and equal to the equilibrium output  $\mathfrak{e}_{\mu}$  of the error system. We will do this by first determining in Section \ref{mme}, a set of values  of $\mu $ for which $z$ and $x$ are  nonconstant for all $t\geq 0.$ Then in Section \ref{horns} we will show that for such values of $\mu$, the distorted but
  infinitesimally  rigid formation $\{\mathbb{G},x(t)\}$ moves  in a circular orbits about the origin in $\R^2$
at a fixed angular speed $\omega_{\mu}$.

 \subsubsection{Mismatch Errors for which $z$ is Nonconstant}\label{mme}

 The aim of this sub-section is to show that once the error $e(\bar{H}x(t))$ has converged to a constant value,
neither $z$ nor $x$ will be constant for small  $||\mu ||$ other than possibly for certain exceptional values.  We shall do this assuming that  the target formation $\{\mathbb{G},y\}$ is ``unaligned'' where
by an {\em unaligned } formation is meant a
formation   $\{\mathbb{G},x\}$ with multi-point
 $x=\matt{x_1' &x_2' & \cdots x_n'}' \in\R^{2n}$ which
does not contain a  set of points $x_i,x_j,x_k,x_l$ in $\R^2$ for which the line between $x_i$ and $x_j$ is parallel to the line between $x_k$ and $x_l$. This is equivalent to saying that $\{\mathbb{G},x\}$ is unaligned if, for every set of     four points $x_i,x_j,x_k,x_l$ within $x$,
$(x_i-x_j)\wedge (x_k-x_l) \neq  0$.  It is clear that  the set of
 multi-points  $x$ for which $\{\mathbb{G},x\}$ is unaligned   is open and dense in $\R^{2n}$.

Throughout this sub-section  we assume that $\scr{A}_{\scr{B}}$ is as in Lemma \ref{xmasday}, that
  $\scr{M}$ is as in Theorem \ref{opp} and that  for each  $\mu\in\scr{M}$, $ x(t,\mu)$ is a solution in $\scr{A}_{\scr{B}}$ to the overall system  for which $e(\bar{H}x(t,\mu)) = \mathfrak{e}_{\mu}$  where $\mathfrak{e}_{\mu}$ is the equilibrium output  of the error system $\dot{\epsilon} = g(\epsilon,\mu)$. Our ultimate goal is to  show that $z$ is nonconstant for  small normed but otherwise ``generic'' values of $\mu $. The following Proposition enables us to make precise what is meant by a generic value.

  \begin{proposition} 
 If  the target formation $\{\mathbb{G},y\}$ is  unaligned, there is an open set  $\scr{M}_0\subset \scr{M}$ about $\mu=0$ within
   which the set of values of $\mu$ for which $z(x(t,\mu))$ is nonconstant  along an overall system  solution $x(t)$ in $\scr{A}_{\scr{B}}$ for which $e(\bar{H}x(t)) = \mathfrak{e}_{\mu}$
   is open and dense in $\scr{M}_0$.
\label{headache} \end{proposition}

What the proposition is saying is that for almost any value of $\mu$ within any sufficiently small open subset of $\R^m$
which contains the origin,  $z(x(t,\mu))$ will be nonconstant along any trajectory of the overall system
in $\scr{A}$ for which  $e$ is fixed
 at the equilibrium output $\mathfrak{e}_{\mu}$ of the error system.
   Thus if $\mu$ has a sufficiently small norm   and otherwise chosen at random, it is almost for certain that   $z(x(t,\mu))$ will be nonconstant. It is natural to say that   $\mu$ is {\em generic}, if it is a  value in $\scr{M}_0$ for which  $z(x(t,\mu))$ is nonconstant.

 The proof of   Proposition \ref{headache} relies on a number of ideas. We begin  with the following construction which
 provides a partial characterization  of the values of $\mu$ for which $z$ is nonconstant.

Let $x=\matt{x_1'&x_2'&\cdots &x_n'}'$ be any multi-point at which $\{\mathbb{G},x\}$ is an infinitesimally rigid formation. As is well known,
the corresponding rigidity matrix $R =R(z)|_{z=\bar{H}x}$ has a kernel of dimension  three \cite{asi.rothII}.
 We'd like to compute an orthogonal basis for $\ker R$.  Towards this end, first recall that rank $H_{m\times n} = n-1$ because $\mathbb{G}$ is a connected graph; thus
$\ker H$ must be a one dimensional subspace and because of this $\ker\bar{H}$ must be of dimension two.
It is well known and  easy to verify that the vectors $q_1 = \matt{ 1 & 0 & 1 &0 &\cdots & 1 & 0}'$  and
 $q_2 = \matt{ 0 & 1 & 0 &1 &\cdots & 0 & 1}'$  constitute an orthogonal basis  for $\ker\bar{H}$.
Next recall that
$R = D'\bar{H}$.  This implies  that $q_1 $ and $q_2$ are in $\ker R$.
It is easy to verify that a  third linearly independent vector
in $\ker R$ is $q_3 = \matt{(Kx_1)' &(Kx_2)' & \cdots &(Kx_n)'}'$
where  $K = \matt{0& -1\cr 1 & 0}$.

To proceed, let
$$v_{\rm avg}(x)  =  \frac{1}{n}\sum_{i=1}^n x_i$$
and define
$q_0 = q_3 + v_2q_1  -v_1q_2$
where $\matt{v_1 & v_2}' = v_{\rm avg}(x)$.
Since span $\{q_0,q_1,q_2\}$ and span
$\{q_1,q_2,q_3\}$ are clearly equal,  the set $\{q_0,q_1,q_2\}$ must be a basis for $\ker R$.
By direct calculation, $q_0'q_i = 0,\;i\in\{1,2\}$, which means that $\{q_0,q_1,q_2\}$
 must be an orthogonal set. Therefore $\{q_0,q_1,q_2\}$ is an orthogonal basis for $\ker R$. With this
  basis in hand we can now give an explicit necessary and sufficient condition for $\dot{z}(\bar{H}x(t,\mu))$ to equal zero at any value of $x$   along a trajectory $x(t,\mu),\;t\geq 0$
  in $\scr{A}$ of the overall system at which $\dot{e}(\bar{H}x(t,\mu))|_{x(t,\mu) = x} = 0$.

\begin{lemma} Let  $\mu \in\R^m$ be fixed
 and for $k\in\mathbf{m}$, let  $(i_k,j_k)$ denote
 the arc from  vertex $i_k$ to  vertex $j_k$ which corresponds
 to edge $k$
in the oriented graph $\mathbb{G}$.  Suppose that
 multi-point  $x=\matt{x'_1 &x'_2&\cdots &x'_n}'$
is a state of the overall system along a trajectory in $\scr{A}$ at which
  $$\dot{e}(\bar{H}x(t,\mu))|_{x(t,\mu) = x} = 0$$
 Then at this value of $x$, $\dot{z}(\bar{H}x(t,\mu))|_{x(t,\mu) = x} = 0$
 if and only if
\eq{w(x)'\mu = 0\label{muer}}
where
\eq{w(x)=\left[
           \begin{array}{c}
             (x_{i_1}-v_{\rm avg}(x))\wedge(x_{j_1}-v_{\rm avg}(x)) \\
             \vdots \\
             (x_{i_m}-v_{\rm avg}(x))\wedge(x_{j_m}-v_{\rm avg}(x)) \\
           \end{array}
         \right].
\label{auddi}}
\label{dd}\end{lemma}
 The reason why this lemma only partially characterizes the values of $\mu$ for which $z$ is nonconstant, is because the state $x$ in \rep{muer} depends on $\mu$. The proof of this lemma is given in the appendix.

\noindent{\bf Triangle Example  Continued:}
Equation \rep{muer} simplifies considerably in the case of a triangular formation. For such a formation with
 coordinate vectors $x_1,x_2,x_3$,  %
 we can always assume \{without loss of generality\}   a graph orientation
 for which  $x_{i_3} = x_1 = x_{j_1}$,
$x_{i_1} = x_2 = x_{j_2}$  and $x_{i_2} = x_3 = x_{j_3}$.  Under these conditions  it is easy to check that for $k\in\{1,2,3\}$,
$$(x_{i_k}-v_{\rm avg}(x))\wedge(x_{j_k}-v_{\rm avg}(x)) =
\frac{1}{3}(x_2-x_1)\wedge(x_1-x_3).$$
Thus for this example,
 $\dot{z}(\bar{H}x(t))|_{x(t) = x} = 0$ if and only if
\eq{(x_2-x_1)\wedge(x_1-x_3)(\mu_1+\mu_2+\mu_3) = 0 \label{triii}}
where $ \matt{\mu_1&\mu_2&\mu_3}' = \mu$.

We now return to the development of ideas needed to prove Proposition \ref{headache}.



\begin{lemma}
Let $\scr{B}$ and $\scr{A}_{\scr{B}}$ be as in the statement of Lemma \ref{xmasday}.  Let $p,q$ be distinct integers in  $\mathbf{m}$.
There is a continuous function $\alpha:\scr{B}\rightarrow \R$ for which
\eq{(x_p-v(x))\wedge(x_q-v(x)) = \alpha(e(\bar{H}x)),  \;\;\;x\in\scr{A}_{\scr{B}} \label{diffe}}
where $v(x)$ is any fixed linear combination of the position vectors $x_i,\; i\in\mathbf{n}$ in $x$.
If, in addition,  the target formation $\{\mathbb{G},y\}$ is unaligned,  then $\alpha (0) \neq 0$. Moreover, if  $\scr{B}$ is sufficiently small, then  $\alpha $   is  continuously differentiable.
\label{resub}\end{lemma}
A proof of this lemma is given in the appendix.



\begin{lemma} Let $\scr{S}\subset\R^m $ be an open subset containing the origin. Let  $f: \scr{S} \rightarrow \R^{1\times m}$ be a continuously differentiable function such that $f(0) \neq 0$. Then there exists an open neighborhood ${\cal U} \subset \scr{S}$ of the origin within which the set of $s$ for which   $f(s)s\neq 0$ is
 open and dense  in $\cal U$.
\label{ali}\end{lemma}
A proof of this lemma is given in the appendix.

\noindent{\bf Proof of Proposition \ref{headache}:}
Let $w(\cdot )$ be as in the statement of Lemma \ref{dd}.
By hypothesis, $\{\mathbb{G},y\}$ is an unaligned formation and $x(t,\mu) \in\scr{A}_{\scr{B}}$ for  all $t\geq 0$ and all $ \mu\in\scr{M}$. In view of
 Lemma \ref{resub},  for $\scr{B}$  sufficiently small the $\;i$th term
 in the row vector $w(x(t,\mu))$
 can   be written as  $\alpha_i(e(\bar{H}x(t,\mu)))$ where  $\alpha_i:\scr{B}\rightarrow \R$ is a continuously differentiable function  satisfying  $\alpha_i(0) \neq 0$. Since this is true for all $m$ terms in $w$, there
 must be a continuously differentiable function $\beta:\scr{B}\rightarrow \R^{1\times m}$  satisfying $\beta(0)\neq 0$ for which $w(x(t,\mu)) = \beta(e(\bar{H}x(t,\mu))),\; t\geq 0, \;\mu\in\scr{M}$. But $e(\bar{H}x(t,\mu)) = \mathfrak{e}_{\mu},\;\;
 t\geq 0,\;\mu\in\scr{M}$ where $\mathfrak{e}_0 = 0$ and  $\mu\longmapsto \mathfrak{e}_{\mu}$ is continuously differentiable. Thus there is a continuously differentiable function  $f:\scr{M}\rightarrow\R^{1\times m}$ for which $f(0)\neq 0$
 and  $w(x(t,\mu)) = f(\mu),\;t\geq 0,\;\mu\in\scr{M}$.

 Note that $\dot{e}(\bar{H}x(t,\mu)) = 0$ because $e(\bar{H}x(t,\mu)) = \mathfrak{e}_{\mu}$. Therefore,  according to Lemma \ref{dd}, $\dot{z}(\bar{H}x(t,\mu)) = 0$  for some $t$ and $\mu\in\scr{M}$ if and only if  $w(x(t,\mu))\mu = 0$. But
 $w(x(t,\mu))\mu = f(\mu)\mu$ for all $t\geq 0$. Therefore $z(x(t,\mu))$ is constant for all $t\geq 0$
  if and only if  $f(\mu)\mu = 0$. But by  Lemma \ref{ali}, there exists a neighborhood
$\scr{M}_0\subset\scr{M}$ of the origin  within which the set of $\mu$ for which $f(\mu)\mu \neq 0$ is open and dense in
$\scr{M}_0$. Thus for any $\mu$ in an open  dense subset of $\scr{M}_0$,  $z(x(t,\mu))$ is nonconstant on $[0,\infty)$
 $\qed $

\subsubsection{Equilibrium Solutions}\label{horns}
The aim of this section is to discuss the evolution of the formation $\{\mathbb{G},x(t,\mu)\}$ along
 an ``equilibrium  solution''   to the overall system assuming that $\mu$ is fixed at any value in $\scr{M}$.
 By an {\em equilibrium solution},  written $\bar{x}(t)$, is meant any solution to \rep{overall} in $\scr{A}$  for which $e(\bar{H}\bar{x}(t)) = \mathfrak{e}_{\mu},\;t\geq 0$,  where $\mathfrak{e}_{\mu}$ is the equilibrium output  of the error system.
   For simplicity we  write  $\bar{z}(t)$ for  = $z(\bar{x}(t))$ and let  
  $\bar{z}_i(t),\;i\in\mathbf{m} $,  be the  sub-vectors in $\R^2$  comprising $\bar{z}(t)$; i.e., $\bar{z}(t) = \matt{\bar{z}_1'(t) &\bar{z}_2'(t) &\cdots &\bar{z}'_n(t)}'$.

Note that since $e(\bar{H}\bar{x}(t)) \in\scr{B}$,  $\bar{x}(t)\in\scr{A}_{\scr{B}}$ and therefore  $\bar{z}(t) \in \bar{H}\scr{A}_{\scr{B}},\;t\geq 0$. Thus, in view of Proposition \ref{resubmit}, there are integers  $p,q\in\mathbf{m}$
 for which the matrix $\bar{Z}(t) = \matt{\bar{z}_p(t) &\bar{z}_q(t)} $ is nonsingular for $t\geq 0$. Moreover
 \eq{\matt{\bar{z}_1 &\bar{z}_2 &\ldots &\bar{z}_m} =\bar{Z}(t) \bar{Q},\;\;t\geq 0\label{sb1n}} and \eq{\dot{\bar{Z}} = \bar{Z}\bar{A},\;\;t\geq 0\label{Z2}}
 where $\bar{Q}$ and $\bar{A}$ are the {\em constant} matrices $\bar{Q} = Q(\mathfrak{e}_{\mu})$ and $\bar{A} = A(\mathfrak{e}_{\mu},\mu)$.
It follows that the Gramian $\bar{Z}'\bar{Z}$ must satisfy
\eq{\dot{\bar{Z}'\bar{Z}} = \bar{A}'\bar{Z}'\bar{Z} +\bar{Z}'\bar{Z}\bar{A}.\label{gr}}

 In view of the definition of the $z_i$ in \rep{pi}, we see that the four entries in $\bar{Z}'\bar{Z}$  are  of the form
 $(\bar{x}_i(t)-\bar{x}_j(t))'(\bar{x}_k(t)-\bar{x}_l(t))$ for various values of $i,j,k$ and $l$.
    But  $\bar{x}(t)\in\scr{A},\;\;t\geq 0$, so as a consequence of   Proposition \ref{quadlemma2}, each such term is equal to a term of the form $\eta_{ijkl}(\mathfrak{e}_{\mu})$ which is constant.  Therefore
 $\bar{Z}'\bar{Z}$ is constant on $[0,\infty)$.
Hence
\eq{ \bar{A}'\bar{Z}'\bar{Z} +\bar{Z}'\bar{Z}\bar{A} = 0 \label{organ}} because of \rep{gr}.
Clearly
$$(\bar{Z}\bar{A}\bar{Z}^{-1})'+ \bar{Z}\bar{A}\bar{Z}^{-1} = 0. $$
Evidently the $2\times 2$  matrix $\bar{Z}\bar{A}\bar{Z}^{-1}$ is skew symmetric so its spectrum must be $\{j\omega, -j\omega \}$  for some real number $\omega \geq 0$.  But $\bar{A}$ is similar to $\bar{Z}\bar{A}\bar{Z}^{-1}$ so $\bar{A}$ must have the same spectrum.

  We claim that $\bar{A} = 0$ and consequently that $\omega = 0$ if and only if $\bar{z}$ is constant.
  To understand why this is so,  note first that  if $\bar{z}$ is constant, then $\dot{\bar{Z}} = 0$. On the other hand, if  $\dot{\bar{Z}} = 0$ then $\bar{z}$ must be constant because of  \rep{sb1n}.  Meanwhile $\dot{\bar{Z}} = 0$ if and only if $\bar{A} = 0$ because of \rep{Z2} and the fact that $\bar{Z}$ is nonsingular. Thus the claim is true.

  Suppose $\bar{z}$ is nonconstant;  as noted in Proposition \ref{headache}, this will be so if $\mu\in\scr{M}_0$.  Then one has $\omega >0$ in which case
  $\bar{z}_p$  and $\bar{z}_q(t) $ must be
   sinusoidal vectors  varying at a single  frequency $\omega$.
Moreover the same must also be true of the remaining $\bar{z}_i$ because of \rep{sb1n}. Additionally, each $\bar{z}_i$ must have a constant norm because for all $t\geq 0$,  $||\bar{z}_i(t)||^2 =  e_i(\bar{H}\bar{x}(t)) +d_i^2,\;i\in\bf{m}$, and $e(\bar{H}\bar{x}(t)) = \mathfrak{e}_{\mu}$. These properties imply that $z_i$ must be  of the form
$$\bar{z}_k(t) = (\bar{\mathfrak{e}}_k +d_k^2)^{\frac{1}{2}}\matt{\;\;\;\cos(\omega t +\phi_k)
 \cr \sigma_k \sin (\omega t+\phi_k )}$$
where $\bar{\mathfrak{e}}_k$ is the $k$th component of $\mathfrak{e}_{\mu}$ and
 $\sigma_k $ equals either $1$ or $-1$.  We claim that  all of the $\sigma _k$ must
be  equal.  To understand why this is so, observe that for all $i,j\in\mathbf{m}$, \small $$\frac{d(\bar{z}'_i\bar{z}_j)}{dt}=\omega (\bar{\mathfrak{e}}_i +d_i^2)^{\frac{1}{2}}(\bar{\mathfrak{e}}_j +d_j^2)^{\frac{1}{2}}(\sigma_i\sigma_j-1)\sin(2\omega t + \phi_i+\phi_j).$$ \normalsize
Since each  $\bar{z}_i'\bar{z}_j$ is constant  and $\omega (\bar{\mathfrak{e}}_i +d_i^2)^{\frac{1}{2}}(\bar{\mathfrak{e}}_j +d_j^2)^{\frac{1}{2}}\neq 0$,
it must be true that $\sigma_i\sigma_j-1  = 0,\;\;i,j\in\mathbf{m}$. Therefore $\sigma_i=\sigma_j, \;\;i,j\in\mathbf{m}$ so all of the $\sigma_k$ have the same value.
We are led to the following result.

\begin{proposition}
Let  $\mu \in\scr{M} $  be fixed and let $y$ be any $\{\mathbb{G},y\}$ be a target formation.
  Suppose the error system  is
in equilibrium  with output  $\mathfrak{e}_{\mu}$.
 Suppose $\bar{x}$ is a solution in $\scr{A}$  to the overall system  along which
  $e(\bar{H}\bar{x}(t))= \mathfrak{e}_{\mu},\;t\geq 0$.
  Then either each $\bar{z}_k(t)$ is constant with norm squared $\bar{\mathfrak{e}}_k+d_k^2$ or there exist
   phase angles $\phi_k,\;\;k\in\mathbf{m}$, and a  frequency $\omega > 0 $
such that
\eq{\bar{z}_k(t) = (\bar{\mathfrak{e}}_k +d_k^2)^{\frac{1}{2}}\matt{\;\;\cos(\omega t +\phi_k)
 \cr \sigma\sin (\omega t+\phi_k )},\;\;k\in\mathbf{m}\label{zequil}}
 where $\bar{\mathfrak{e}}_k$ is the $k$th component of the equilibrium output $\mathfrak{e}_{\mu}$ and $\sigma $ is a constant with value $1$ or $-1$.
\label{main}\end{proposition}

\noindent It is worth noting that if $\sigma =1$ then all of the $\bar{z}_k$ rotate about the origin in $\R^2$ in a clockwise direction, while if $\sigma  = -1$, the $z_i$ all rotate in a counter-clockwise direction.

We  are now in a position to more fully   characterize any equilibrium solution $\bar{x}$.
Two situations can occur:  Either $\bar{z}(t)$ is constant or it is not.  We  first consider the case when $\bar{z}(t)$
  is constant.
  Examination of \rep{overall} reveals that if $\bar{z}$ is constant then so is $\dot{\bar{x}}$. This means  that the any  formation $\{\mathbb{G},\bar{x}\}$  for which $\bar{z}$ is constant is
  either  stationary   or  it drifts to off infinity at a constant velocity, depending on the value of $\mu\in\scr{M}$. If there is no mismatch \{ie, $\mu =0$\}, then $\mathfrak{e}_{\mu} = 0$, as noted just below  Corollary \ref{frunch}. In this  case $\bar{x}$ must therefore  be constant   and the formation must be stationary and  have   desired shape. The following example illustrates
  that formations for which   $\bar{z}$ is constant, can in fact  drift off to infinity for some values of $\mu\in\scr{M}$.

\noindent{\bf Triangle Example  Continued:}
 We claim that under the conditions that $\bar{z}$ is constant and $\mu\neq 0$, the velocity of the average vector
 $v_{\rm avg}  =  \frac{1}{3}( \bar{x}_1+\bar{x}_2 +\bar{x}_3)$ is  a nonzero constant which means that with mismatch, the triangular
 formation  $\{\mathbb{G},\bar{x}\}$ must drift off to infinity at a constant velocity. To understand why this is true,  note that  the mismatch errors must satisfy the non-generic condition
 \eq{\mu_1 +\mu_2+\mu_3 = 0\label{ppoiu}}
 because of Lemma \ref{dd}, \rep{triii} and the hypothesis that $\bar{z}$ is constant. Meanwhile from \rep{epdates0},
 $\dot{v}_{\rm avg} = \bar{z}_1\mu_1 +\bar{z}_2\mu_2+\bar{z}_3\mu_3$, so if $\dot{v}_{\rm avg}$ were zero, then
 $\bar{z}_1\mu_1 +\bar{z}_2\mu_2+\bar{z}_3\mu_3=0$. But for the triangle, $\bar{z}_1+\bar{z}_2+\bar{z}_3=0$ which means that $\bar{z}_1(\mu_1-\mu_3) +\bar{z}_2(\mu_2-\mu_3)=0$.  However $\bar{z}_1$ and $\bar{z_2}$ are linearly independent because
 the formation $\{\mathbb{G},\bar{x}\}$ is infinitesimally rigid. Therefore the coefficients
 $\mu_1-\mu_3$ and $\mu_2-\mu_3$ must both be zero which means that $\mu_1=\mu_2 =\mu_3$. This and \rep{ppoiu} imply that
 $\mu = 0$ which contradicts  the hypothesis that $\mu\neq 0$. Thus for the triangular formation,   $\dot{v}_{\rm avg}$ is a nonzero constant as claimed.\vspace{.3in}

We now turn next to the case when $\bar{z}$ is nonconstant. In view of Proposition \ref{headache},  this case is anything but vacuous. We already know that in this case, $\omega >0$.
In view of \rep{ov}, and the assumption that $e(\bar{H}\bar{x}(t)) = \mathfrak{e}_{\mu}$, it is clear that  the $\bar{x}_i$ satisfy the differential equations
\eq{\dot{\bar{x}}_i =  -\sum_{j\in\scr{N}_i^+}\bar{z}_{k_{ij}}\bar{\mathfrak{e}}_{k_{ij}}
 + \sum_{j\in\scr{N}_i^-}\bar{z}_{k_{ij}}(\bar{\mathfrak{e}}_{k_{ij}}+\mu_{k_{ij}}),\;\;\;\;i\in\mathbf{n}. \label{ovuu}}
Note that the   right hand sides of these differential equations are  sinusoidal  signals at frequency $\omega $
because the $\bar{\mathfrak{e}}_i$ and $\mu_i$ are constants.
This means that the $\bar{x}_i$  must be  of the form
  $$\bar{x}_i(t) = \matt{\;\;a_i\cos (\omega t +\theta_i) \cr \sigma b_i \sin (\omega t +\gamma_i)}  + q_i,\;\;i\in\mathbf{n}$$
  where the $q_i$ are constant vectors in $\R^2$ and the $a_i$, $b_i$, and $\theta_i$ are real numbers with $a_i>0$.
Note, in addition, from \rep{ovuu} that  for each $i$, $||\dot{\bar{x}}_i||^2$ can be written as a linear combination of terms of the form $\bar{z}_j'\bar{z}_k$ for various values of $j$ and $k$. But in view of \rep{pi} and Proposition \ref{quadlemma2},
 each such term $\bar{z}_j'\bar{z}_k$  is a function of $e(\bar{H}\bar{x}(t))$ which in turn equals $\mathfrak{e}_{\mu}$ which  is a constant.  Thus each norm
  $||\dot{\bar{x}}_i||$ must be a finite constant. This means that $\gamma_i =\theta_i$, $b_i=a_i$  and thus that  each $\bar{x}_i$ is of the form
   \eq{\bar{x}_i(t) = a_i\matt{\;\;\cos (\omega t +\theta_i) \cr \sigma \sin (\omega t +\theta_i)}  + q_i,\;\;i\in\mathbf{n}.\label{pppop}}
We claim that all of the  $q_i$ are equal to each other. That is, there is a single vector $q$ for which
\eq{\bar{x}_i(t) = a_i\matt{\;\;\cos (\omega t +\theta_i) \cr \sigma\sin (\omega t +\theta_i)}  + q,\;\;i\in\mathbf{n}.\label{rotate}}
To understand why this is so, note first that \rep{zequil} implies that
$\dot{\bar{z}}_k = \omega K\bar{z}_k,\;\;i\in\mathbf{n}$, where $K = \matt{0 & -1\cr 1 & 0}$.  Suppose that $\bar{x}_i$ and $\bar{x}_j$ are the coordinate vectors for which $\bar{z}_k = \bar{x}_i-\bar{x}_j$. Then
\eq{\dot{\bar{x}}_i -\dot{\bar{x}}_j = \omega K (\bar{x}_i -\bar{x}_j).\label{uio}}
But from \rep{pppop},
$$\dot{\bar{x}}_i = \omega K(\bar{x}_i - q_i),\;\;\;i\in\mathbf{n}$$
so
$$\dot{\bar{x}}_i -\dot{\bar{x}}_j = \omega K(\bar{x}_i - \bar{x}_j)) + \omega K(q_i - q_j).$$
From this and \rep{uio} it follows that $\omega K(q_i - q_j) = 0$ and thus that $q_i=q_j$. Since this argument applies to
all edges in a connected graph $\mathbb{G}$, it must be true that all $q_i$ are equal as claimed.
We are led to the following characterization of equilibrium solutions.

\begin{theorem} Let $\mu\in\scr{M}$ be fixed and suppose that $\{\mathbb{G},y\}$  is a target formation. Let    $\bar{x}$ be  a solution  in $\scr{A}$ to the overall system
  along which  $e(\bar{H}\bar{x}(t))=\mathfrak{e}_{\mu}$.
 \begin{enumerate}
 \item  If $\mu$ is a mismatch error for which $\bar{z}$ is constant, then depending on the value of $\mu$,  all points  within the  time-varying, infinitesimally  rigid formation  $\{\mathbb{G},\bar{x}(t)\}$ with distorted edge distances $(\bar{\mathfrak{e}}_i + d_i^2)^{\frac{1}{2}},\;i\in\mathbf{m}$, are either fixed in position or move off to infinity at the same constant velocity.
 \item If  $\mu$ is a mismatch error  for which $\bar{z}$ is nonconstant,
then
  all points  within  $\{\mathbb{G},\bar{x}(t)\}$
   rotate in either a clockwise or counterclockwise direction with the same constant angular speed $\omega >0 $ along  circles  centered at  some point $q$ in the plane, as does the distorted
   formation itself.   Moreover if the target formation $\{\mathbb{G},y\}$ is unaligned, almost any mismatch error $\mu$ will cause this behavior to occur provided
    the norm of  $\mu $ is sufficiently small.\end{enumerate}\label{main2}\end{theorem}


\subsection{Non-Equilibrium Analysis}\label{NEA}

Fix $\mu\in \scr{M}$.
 In this section we will consider the situation when a solution $x(t)$ of the overall system starts out with an error signal $e(\bar{H}x(t))$ which is initially close to the equilibrium output $\mathfrak{e}_{\mu}$  of the error system. As in section \ref{horns}, we let $\bar{x}(t)$ denote an equilibrium solution of the over all system  and we write $\bar{z}(t) = z(\bar{H}\bar{x}(t))$.
There may of course be many equilibrium solutions $\bar{x}(t)$ to the overall system  along which
$e=\mathfrak{e}_{\mu}$. 
 Our aim is to show that any solution  to the overall system
  starting with $e(\bar{H}x(0))$ sufficiently close to $\mathfrak{e}_{\mu}$
converges exponentially fast to  such  an   equilibrium solution.

  Let
$\dot {\epsilon} = g(\epsilon, \mu)$ be the error system  and let $\scr{A}\subset\scr{X}$ be an ambient space
 on which it is valid. Let $\scr{B}$ be any ball satisfying the hypotheses of Lemma \ref{xmasday}.
  We know already from Theorem \ref{opp} that with $||e(\bar{H}x(0) - \mathfrak{e}_{\mu}||$ sufficiently small with $x(0)\in\scr{A}$,  $x(t)$ exists and is in $\scr{A}_{\scr{B}}$ for all time  and $e(\bar{H}x(t))$  converges exponentially fast to
  $\mathfrak{e}_{\mu}$. We assume that $||e(\bar{H}x(0) - \mathfrak{e}_{\mu}||$ is this small.
We also know from Proposition \ref{resubmit} that there are integers $p,q\in\mathbf{m}$ and time-varying  matrices $Q(e(\bar{H}x(t)))$
and $A(e(\bar{H}x(t)),\mu)$, henceforth denoted by $Q(t)$ and $A(t)$ respectively, for which
\eq{\matt{z_1 &z_2 &\ldots &z_m} =Z Q,\;\;\label{nnsb1}}
and
\eq{\dot{Z} = ZA\label{nnsq1}}
where $z_i\in\R^2$ is the $i$th component sub-vector  of $z=\bar{H}x(t)$ and $Z$ is the nonsingular, time-varying matrix
$Z = \matt{z_p &z_q}$. Since $e(\bar{H}x(t))$ converges to $\mathfrak{e}_{\mu}$ exponentially
fast, $Q$ and $A$ converge exponentially fast to constant matrices $\bar{Q} = Q(\mathfrak{e}_{\mu})$ and  $\bar{A} = A(\mathfrak{e}_{\mu},\mu)$ respectively. Note that  because of \rep{e}, $||z_i||^2 = d_i^2 + e_i,\;i\in\mathbf{m}$, where $e_i$ is the $i$th component of $e(\bar{H}x(t))$. Thus for $i\in\mathbf{m}$,   $||z_i||^2$  converges to $d_i^2+ \bar{\mathfrak{e}}_i$ where $\bar{\mathfrak{e}}_i$ is the $i$th component of $\mathfrak{e}_{\mu}$.
Therefore the  $z_i$ and $Z$ must be bounded on $[0,\infty)$.
 Note in addition that $e^{\bar{A}t}$ must be periodic and consequently bounded  on the whole real line $(-\infty,\infty)$ because  either $\bar{A} = 0$, or if it is not, its spectrum must be  $\{j\omega,-j\omega\}$ for some $\omega >0$.

Let $V_{2\times 2}$ be that solution to $\dot{V} = V\bar{A}$ with initial state
\eq{V(0) = Z(0) + \int_0^{\infty}U(\tau)e^{-\bar{A}\tau}d\tau\label{po}}
where $U =Z(A-\bar{A})$. Note that $U$  tends to zero
  exponentially fast
  because
$Z$ is bounded and because
 $A-\bar{A}$ tends to zero exponentially fast.
Observe that $V(0)$ exists because $e^{-\bar{A}t}$
 is bounded on $[0,\infty)$ and because $U$ tends to zero exponentially fast. Note that
  $V$ must be periodic because $e^{\bar{A}t}$ is.
We claim that $Z$ converges exponentially fast  to $V$  as $t\rightarrow\infty$.  To understand why this is so,
 consider the error $E = Z-V$ and note  that
$$\dot{E} = E\bar{A} + U.$$
By  the variation of constants formula
$$E(t) = E(0)e^{\bar{A}t} +\int_0^tU(\tau)e^{\bar{A}(t-\tau)}d\tau.$$
In view of \rep{po},
$$E(t) = -\int_t^{\infty}U(\tau)e^{\bar{A}(t-\tau)}d\tau.$$
Now since $e^{\bar{A}(t-\tau)}$ is bounded for all $t$ and $\tau$ and $U(\tau)$ tends to zero exponentially fast,
there must exist positive constants $c$ and $\lambda$ such that
$||U(\tau)e^{\bar{A}(t-\tau)}||\leq ce^{-\lambda \tau}$.
Clearly $||E(t)||\leq \int_t^{\infty}ce^{-\lambda\tau}d\tau = \frac{c}{\lambda}e^{-\lambda t} $
so $E(t) \rightarrow 0$ as $t\rightarrow\infty$ as fast as $e^{-\lambda t} $ does. It follows that
$Z$ converges exponentially fast to $V$  as claimed.

Let $v_p$ and $v_q$ denote the columns of $V$ and for all $i\in\mathbf{m}$ except for $i\in\{p,q\}$, define $v_i = V\bar{Q}\zeta_i$  where $\zeta_i$  is the $i$th  unit vector in $\R^m$. We claim that for  $i\in\mathbf{m}$,
  $z_i$ converges to $v_i$ exponentially fast. To understand why this is so, note that because of $Q$'s its definition in the proof of Proposition \ref{resubmit},
  $Q\zeta_p =  \nu_1$  and  $Q\zeta_q =  \nu_2$ where $\nu_i$ is the $i$th unit vector in $\R^2$. Since $Q$ converges to $\bar{Q}$,  $\bar{Q}\zeta_p =  \nu_1$  and  $\bar{Q}\zeta_q =  \nu_2$.
From this it follows that $v_p= V\bar{Q}\zeta_p$, $v_q=V\bar{Q}\zeta_q$, and thus that
 $v_i = V\bar{Q}\zeta_i,\;i\in\mathbf{m}$. Hence, for each $i\in\mathbf{m}$,
 $z_i-v_i = ZQ\zeta_i - V\bar{Q}\zeta_i$. Therefore for each such $i$,
$z_i-v_i = (Z(Q-\bar{Q}) +(Z-V)\bar{Q})\zeta_i$.
But $Z$ and $V$ are bounded signals and $Z\rightarrow V$ and $Q\rightarrow \bar{Q}$ so clearly for $i\in\mathbf{m}$,
  $z_i$ converges to $v_i$ exponentially fast as claimed.

 We now claim that
\eq{||v_i(t)||^2 = \bar{\mathfrak{e}}_i +d^2_i,\;t\geq 0, \;i\in\mathbf{m}\label{last1}}  where, as before, $\bar{\mathfrak{e}}_i$ is the $i$th component of $\mathfrak{e}_{\mu}$.
To understand why this is so, recall that $||z_i||^2 = e_i(\bar{H}x(t)) +d_i^2$ because of \rep{e}.  Moreover $e_i(\bar{H}x(t))$ converges exponentially fast
  to $\bar{\mathfrak{e}}_i$.
 Thus $||z_i||^2$ converges to  $\bar{\mathfrak{e}}_i +d_i^2$.  We know that $||z_i||^2$ converges to $ ||v_i||^2$ because  $z_i$ converges to $v_i$. Therefore $||v_i||^2$ converges to $\bar{\mathfrak{e}}_i +d_i^2$. But each $v_i$ is a sinusoidally varying vector at frequency $\omega$   because $V$ is a solution to $\dot{V}=V\bar{A}$. This means that each norm $||v_i||^2$ is periodic. Thus the only way  $||v_i||^2$  can converge is if it is constant to begin with.
Therefore
$||v_i(t)||^2 = \bar{\mathfrak{e}}_i +d^2_i$ for all $t\geq 0$ as claimed.

To conclude we need to construct an equilibrium solution $\bar{x}(t)$ to the overall system to which $x$ converges.
As a first step let us note that the differential equation describing the overall system \rep{epdates0}
can be written as
$\dot{x} = B(e(z),\mu)z$ where $B(e,\mu)$  is  continuous in $e$ and $z=\bar{H}x$. Define
$$\bar{x}(t) =x(0)+\int_0^{\infty}w(\tau)d\tau
 + \int_0^tB(\mathfrak{e}_{\mu},\mu)v(\tau)d\tau $$
 where $w(\tau) =B(e(\bar{H}x(\tau)),\mu)z(\tau)-B(\mathfrak{e}_{\mu},\mu)v(\tau)$  and $v = \matt{v_1'&v_2'\cdots &v_m'}'$.
The integral $\int_0^{\infty}w(\tau)d\tau$ is well defined and finite because $z-v$ and $B(e(\bar{H}x(t)),\mu)-B(\mathfrak{e}_{\mu},\mu)$
  tend to zero exponentially fast.
Our goals are to show that  $x-\bar{x}$ converges to zero exponentially fast and also that $\bar{x}(t)$
is an equilibrium solution to the overall system along which $e(\bar{H}\bar{x}(t))=\mathfrak{e}_{\mu}$. To deal with the first issue observe because of its definition,
\eq{\dot{\bar{x}} = B(\mathfrak{e}_{\mu},\mu)v.\label{horn}} Thus
  the error  vector $q=x-\bar{x}$  satisfies
 $\dot{q} = w(t)$.  Therefore  $q(t) = x(0)-\bar{x}(0) +\int_0^tw(\tau)d\tau $ so $q= -\int_t^{\infty}w(\tau)d\tau$.
 Recall that $w$ converges to zero exponentially fast; therefore by the same reasoning
 which was used to show that $E(t)$ converges to zero exponentially fast, one concludes   that $q$ must converge
  to zero exponentially fast.  Thus $x$ converges to $\bar{x}$ exponentially fast.

  It remains to be shown that $\bar{x}$ is an equilibrium solution.  As a first step towards this end, note that
  $e(v) = \mathfrak{e}_{\mu}$ because  of \rep{last1}. Next note that $\matt{v_1 &v_2 &\ldots & v_m} = V\bar{Q}$ because
  $v_i = V\bar{Q}\zeta_i,\;i\in\mathbf{m}$. But $\dot{V} = V\bar{A}$. Thus $\matt{\dot{v}_1 &\dot{v}_2 &\ldots & \dot{v}_m} = V\bar{A}\bar{Q}$. From this and \rep{pupps} it follows that  $\matt{\dot{v}_1 &\dot{v}_2 &\ldots & \dot{v}_m} =
\matt{v_1 &v_2 &\ldots & v_m}M(\mathfrak{e}_{\mu},\mu)$.  Since $e(v) = \mathfrak{e}_{\mu}$, $v$ must therefore satisfy
\rep{z2}.  But \rep{z2} can also be written as  $\dot{z} =\bar{H}B(e(z),\mu)z$, so $\dot{v} =\bar{H}B(e(v),\mu)v$
or $\dot{v} =\bar{H}B(\mathfrak{e}_{\mu},\mu)v$.  Clearly $\bar{H}\dot{\bar{x}} =\bar{H}B(\mathfrak{e}_{\mu},\mu)v$ because of
\rep{horn}. Hence  $\bar{H}\dot{\bar{x}}= \dot{v}$  so $v = \bar{H}\bar{x} +p$ for some constant vector $p$.

We claim  that $p=0$ and thus that $v = \bar{H}\bar{x}$. To understand why this is so,  recall that each $z_i-v_i,\;i\in\mathbf{m}$ converges to zero, so $v$ converges to $z$. We have also shown that $\bar{x} - x$ converges to zero, so $\bar{H}\bar{x}$ must converge to $z$ which equals $ \bar{H}x$.  Therefore $v-\bar{H}\bar{x}$ must converge to zero and the only why this can happen is if $p=0$. Therefore $v=\bar{H}\bar{x}$. If follows from this and \rep{horn} that
$\dot{\bar{x}} = B(e(\bar{H}\bar{x}),\mu)\bar{H}\bar{x}$. Therefore $\bar{x}$ satisfies \rep{epdates0} with
$e(\bar{H}\bar{x}) = \mathfrak{e}_{\mu}$, so $\bar{x}$ is an equilibrium solution of the overall system.
We are led to
 the following theorem which is the main result
     of this paper.

\begin{theorem} Let $\mu\in\scr{M}$ be fixed.
 Let $x(t)$ be any solution of the overall system
  starting in a state in $\scr{A}$ for which the  reduced error  $\tilde{P}e(\bar{H}x(0))$  is in the domain of attraction of the
   exponentially stable
  equilibrium state $\epsilon_{\mu}$  of the  error system $\dot{\epsilon} = g(\epsilon,\mu)$.
  There exists a  solution    $\bar{x}$
  to the overall system
  along which  $e(\bar{H}\bar{x}(t)=\mathfrak{e}_{\mu}$, to which $x(t)$ converges exponentially fast.
 \begin{enumerate}
 \item  If $\mu$ is a mismatch error for which $\bar{z}$ is constant, then depending on the value of $\mu$,
  all points  within the  time-varying, infinitesimally  rigid formation  $\{\mathbb{G},x(t)\}$ either converge exponentially fast  to constant values or drift off to infinity.

  \item If  $\mu$ is a mismatch error  for which $\bar{z}$ is nonconstant,
then
  all points  within  $\{\mathbb{G},x(t)\}$ converge exponentially fast  to the points in a formation which
rotates in either a clockwise or counterclockwise direction   at a constant angular speed $\omega >0$  along a circle centered at some fixed point in the plane.  Moreover if the target formation $\{\mathbb{G},y\}$ is unaligned, almost any mismatch error $\mu$ will cause this behavior to occur provided
    the norm of  $\mu $ is sufficiently small.\end{enumerate}

   \label{Maino} \end{theorem}

\section{Concluding Remarks}
In this paper we have identified a  basic robustness problem with the type of formation control proposed in  \cite{krickb}.
 A natural question to ask is  if  the problematic behavior   can be eliminated by modifying the control laws?   Simulations suggest that introducing delays or dead zones will not help.  On the other hand,   progress has been  made to achieve robustness by  introducing controls which  estimate the mismatch error and take appropriate corrective action
similar in spirit to what is typically done in adaptive control \cite{HMB13IFAC}.  While results exploiting
this  idea are limited in scope \cite{ACC14.1,cdc14.4}, they do nonetheless suggest that the approach may indeed resolve the problem.

 We see no roadblocks to extending the findings of this paper  to three dimensional formations.  All of the material in Sections \ref{uf} through \ref{redu} is readily  generalizable without any surprising changes, although the square subsystem in Section \ref{redu} will of course have to be $3\times 3$ rather than $2\times2$.  This change in size has an important consequence. This implication is that the skew symmetric matrix $\bar{Z}\bar{A}\bar{Z}^{-1}$ used in Section \ref{horns} to characterize the spectrum of $\bar{A}$, will be $3\times 3$ rather than $2\times 2$.  Thus in the three dimensional case, if
 $\bar{Z}\bar{A}\bar{Z}^{-1}$  is nonzero, its spectrum and consequently  $\bar{A}$'s, must contain an a eigenvalue at $0$ in addition to a pair of imaginary
 numbers $j\omega$ and $-j\omega$. Thus the corresponding formation  will not only
 rotate at an angular speed $\omega$,  but it will also drift linearly with time. More precisely, in the three dimensional case, a mismatch errors can cause formation to  move off   to infinity   along a helical trajectory.
  These observations will be fully justified in a forthcoming paper devoted  to the three dimensional version of the problem.

Other questions remain. For example,
it is natural to wonder how these  findings might change for formations with more realistic dynamic agent models.
We conjecture that more elaborate agent models will not significantly alter the findings of this paper, although actually proving this will likely be challenging, especially in the realistic case when the parameters in the models of different agents are not identical.

Another issue to be resolved is whether or not a formation needs to be unaligned for the last statement of Theorem \ref{Maino} to hold. We conjecture that the assumption is actually not necessary.

Finally we point out that  robustness issues raised here have broader implications extending well beyond formation maintenance   to the entire field of distributed optimization and  control.
In particular, this research  illustrates that when assessing the efficacy of a particular distributed algorithm,
one must consider the consequences  of  distinct agents having
 slightly different understandings of what the values of  shared data   between them is suppose to be. For  without the
 protection of  exponential stability,  it is likely that such discrepancies will cause significant  misbehavior  to occur.


\section{Appendix}

\noindent{\bf Proof of Lemma \ref{china2}:} Suppose that the conclusion of the lemma is false in which case, for each
$k\in\mathbf{n}$, the vectors $x_i-x_k,\;i\in\scr{N}_k$,
span a subspace of  dimensional at most one. Since $\mathbb{G}$ is connected, this means that the set of
 all vectors $x_i-x_j$ for which $(i,j)$ is an edge in $\mathbb{G}$, must also span a subspace of dimensional of at most one, as must the set of all $z_i,\;i\in\mathbf{m}$.
 Thus there must be a vector $w\in\R^2$ and $m$
real numbers $c_i,\;i\in\mathbf{m}$ such that $z_i = c_i w,\;i\in\mathbf{m}$. Therefore $D(z) =C\otimes w$
where $D(z) = {\rm diagonal}\;\{z_1,z_2,\ldots,z_{m}\}_{2m\times m}$ and   $C = {\rm diagonal}\;\{c_1,c_2,\ldots,c_{m}\}$. Then the rigidity matrix for $\{\mathbb{G},x\}$  is $R(z)|_{z=\bar{H}x}$ where $R(z) = D'(z)(H\otimes I_{2\times 2})$ and $H'$ is the incidence matrix of $\mathbb{G}$. Therefore $$R = (C\otimes w)'(H\otimes I_{2\times 2})= (C'\otimes  w')(H\otimes I_{2\times 2})=(C'H)\otimes w'.$$
Thus ${\rm rank} \;R = ({\rm rank}\;C'H)({\rm rank}\;w') \leq {\rm rank}\;C'H$; therefore  ${\rm rank} \;R \leq {\rm rank}\;H$.
  But ${\rm rank}\;H = n-1$ because $H'$ is the  incidence matrix of an $n$ vertex connected graph. Therefore  ${\rm rank} \;R\leq n-1$. Since $n\geq 3$,  this contradicts to the requirement that ${\rm rank }\; R = 2n-3$ which is a consequence of the hypothesis that  $\{\mathbb{G},x\}$ is infinitesimally rigid. $\qed$

\noindent{\bf Proof of Lemma \ref{dd}:}
It will first be shown that  $\dot{z} = 0$  if and only if
\eq{q_0'S'\mu = 0. \label{fri}}
To prove that this is so,
let $U$ and $V$ be full rank matrices such that $R=UV$. Thus $U$  and $V'$ have
 linearly independent columns.  This implies that $\ker V = \ker R$ and that the matrix $VV'$ is nonsingular.
Since $e\dfb e(\bar{H}x)$ is constant, \rep{error1} implies that
$RR'e = RS'\mu$; thus
 $UVV'U'e = UVS'\mu$.  Therefore  \eq{U'e = (VV')^{-1}VS'\mu.\label{tpp}}

In view of \rep{z2},  the condition   $\dot{z} = 0$ is equivalent to
 $\bar{H}(R'e-S'\mu) = 0$  which can be re-written as $\bar{H}(V'U'e-S'\mu) = 0$.
This and \rep{tpp} enable us to write
 \eq{\bar{H}PS'\mu = 0\label{ppp}}
 where
 $$P = V'(VV')^{-1}V - I.$$
Note that $P$ is the orthogonal projection on the orthogonal complement of the column span of $V$ which
is the same as $\ker V$. Since $\ker V = \ker R$,  $P$ is therefore the orthogonal projection on $\ker R$.

Since $\R^{2n} = \ker R \oplus (\ker R)^{\bot}$,  the vector $S'\mu$ can be written as
$S'\mu = \lambda_0q_0 +\lambda_1q_1+\lambda_2q_2 +q_4$ where the $\lambda_i$ are scalars and
 $q_4$ is in the orthogonal complement of $\ker R$. Thus $PS'\mu = \lambda_0q_0 +\lambda_1q_1+\lambda_2q_2$.
 Therefore  $\bar{H}PS'\mu = \lambda_0\bar{H}q_0$ because $q_1$ and $q_2$ are in $\ker \bar{H}$.
Therefore \rep{ppp} is equivalent to  $\lambda_0\bar{H}q_0 =0$; but $\bar{H}q_0 \neq 0$
because $q_0$ is orthogonal to $q_1$ and $q_2$ and $\ker \bar{H} =$ span $\{q_1,q_2\}$.  Therefore $\lambda_0 = 0$ or equivalently, $S'\mu$ must be
 orthogonal to $q_0$.  Therefore  $\dot{z} = 0$ and \rep{fri} are equivalent statements.

Note that $q_0$ can be rewritten as $$q_0=\left[
\begin{array}{c}
 K(x_1-v_{\rm avg}(x)) \\
 \vdots \\
 K(x_n-v_{\rm avg}(x)) \\
 \end{array}
 \right].
$$ Note in addition that for $k\in\mathbf{m}$,
 row vector $x'_{i_k}-x'_{j_k}$ must appear in the $k$th row and
  $i_k$th block column of $S$ and all other terms
 in  row $k$
 of $S$  must be zero. Thus
   the $k$th row of the vector $Sq_0$ must be
$(x_{i_k}-x_{j_k})'K(x_{i_k}-v_{\rm avg}(x))$. This in turn can be written more concisely as
$-(x_{i_k}-v_{\rm avg}(x))\wedge(x_{j_k}-v_{\rm avg}(x))$. It follows from this and the equivalence of $\dot{z} = 0$ and
\rep{fri} that the lemma is true. $\qed$

\noindent{\bf Proof of Lemma \ref{resub}:}
Since a wedge product is a bilinear map and $v(x)$ is a linear combination of the position vectors $x_i,\; i\in\mathbf{n}$, the wedge product $(x_p-v(x))\wedge(x_q-v(x))$
can be expanded and written as a linear combination  of the wedge products $(x_p-x_i)\wedge(x_q-x_j),\;i,j\in\mathbf{n}$.
That is  \eq{(x_p-v(x))\wedge(x_q-v(x)) = \sum_{i,j\in\mathbf{n}}\lambda_{ij}((x_p-x_i)\wedge(x_q-x_j)),\label{lincomb}}
where each $\lambda_{ij}\in\R$ and $x\in\scr{A}_{\scr{B}}$.
But $\scr{A}_{\scr{B}}\subset \scr{A}$. Therefore, as a consequence of  Proposition \ref{quadlemma2},  there are
smooth functions
$f_{ij}:e(\bar{H}\scr{A})\rightarrow\R$ such that \eq{(x_p-x_i)\wedge(x_q-x_j)^2 = f_{ij}(e(\bar{H}x)),\;x\in\scr{A}_{\scr{B}},\;\;i,j\in\mathbf{n}.\label{hp}}
Thus
 the function  $\alpha:\scr{B}\rightarrow \R$,\textbf{\textbf{}}  $e\longmapsto \sum_{i,j\in\mathbf{n}}\lambda_{ij}\sqrt{f_{ij}(e)}$ satisfies \rep{diffe} and is continuous.

Now suppose that $\{\mathbb{G},y\}$ is unaligned. Then $(y_p-y_i)\wedge(y_q-y_j) \neq 0, \;i,j\in\mathbf{n}$.
  Since  $y\in\scr{A}_{\scr{B}}  $, \rep{hp} holds with $x=y$. Moreover, $e(\bar{H}y) = 0$. Therefore
 \eq{ f_{ij}(0)\neq 0,\;\;i,j\in\mathbf{n}.\label{hp2}}
 Hence  $\alpha(0) \neq 0$.

 From \rep{hp2} it is clear that if
 $\scr{B}$  is  small enough,  $f_{ij}(e)\neq 0,\;e\in\scr{B},\;i,j\in\mathbf{n}  $.
Under this condition, the functions $e\longmapsto \sqrt{f_{ij}(e)},\;i,j\in\mathbf{n}$ are all continuously differentiable and so therefore is   $\alpha$.  $\qed $

\noindent{\bf Proof of Lemma \ref{ali}:} The function $h:\scr{S}\rightarrow \R$ defined by $s\longmapsto f(s)s$ is continuously differentiable and $h(0) = 0$.
Moreover,
$$
\frac{\partial h(s)}{\partial s}  =
\frac{\partial f(s)}{\partial s} s + f(s)$$
so
\begin{equation}
\label{eq:diffatzero}
\left. \frac{\partial h(s)}{\partial s}\right |_{s=0}  = f(0) \neq 0.
\end{equation}
This and the fact that $h$ is continuously differentiable imply that  exists a neighborhood $\cal U$ of the origin on which  $\frac{\partial h(s)}{\partial s}$ is non-zero. Since $\frac{\partial h(s)}{\partial s}$ is a nonzero, $1\times m$ matrix, it is therefore of full rank on $\scr{U}$. Therefore every $s\in\scr{U}$ is a regular point of $h(s)$. Hence  $0$  is a regular value of $h(s)$
on $\scr{U}$. Therefore by the
 regular value theorem \cite{Lee09AMS},  the set $$\scr{T} = \left\{ s :s\in {\cal U}, \; h(s)=0 \right\}$$ is a regular submanifold of $\cal U$ of dimension $m-1$. Hence, there exists a neighborhood ${\cal V} \subset \mathbb{R}^{m-1}$ of the origin and a continuously differentiable function  $\phi: \scr{V} \rightarrow \scr{S}$ such that $\phi({\cal V}) = \scr{T}$. By Sard's theorem, which states that the image of $\phi$ has Lebesgue measure zero in $\mathbb{R}^m$, we conclude that
$\scr{T}$ is of measure zero and thus its complement is dense in $\scr{U}$ .
$\qed $

\bibliographystyle{unsrt}
\bibliography{my,steve,mous}

\end{document}